\pdfoutput=1 

\documentclass[twocolumn]{aastex62}

\newcommand\codename{\textsc{monk}}
\defcitealias{dovciak_minimum_2016}{DD16}
\usepackage[T1]{fontenc}
\usepackage[utf8]{inputenc}

\shorttitle{GR Comptonisation}
\shortauthors{Zhang, Dov\v ciak, \& Bursa}

\usepackage{amsmath}
\begin{document}
\title{Constraining the size of the corona with fully relativistic calculations of spectra of extended corona. I - the Monte Carlo
radiative transfer code}

\correspondingauthor{Wenda Zhang}
\email{zhang@asu.cas.cz}

\author{Wenda Zhang}
\affiliation{Astronomical Institute, Czech Academy of Sciences, 
  Bo\v cn\'i II 1401, CZ-14100 Prague, Czech Republic}
\author{Michal Dov\v ciak}
\affiliation{Astronomical Institute, Czech Academy of Sciences, 
  Bo\v cn\'i II 1401, CZ-14100 Prague, Czech Republic}
\author{Michal Bursa}
\affiliation{Astronomical Institute, Czech Academy of Sciences, 
  Bo\v cn\'i II 1401, CZ-14100 Prague, Czech Republic}



\begin{abstract}
The size and geometry of the X-ray emitting corona in AGNs are still not well constrained. \citet{dovciak_minimum_2016} proposed a method 
based on calculations assuming
a point-like lamp-post corona. To perform more self-consistent calculations of energy spectra of extended coronae, 
we develop \codename, a Monte Carlo radiative transfer code dedicated to calculations of Comptonised spectra
in the Kerr spacetime.
In \codename{} we assume Klein-Nishina scattering cross section and include all general relativistic effects.
We find that for a corona located above the disc, the spectrum is not isotropic, but with harder and less luminous spectra
towards observers at lower inclinations, owing to anisotropic illumination of the seed photons.
This anisotropy also leads to an underestimated size of the corona if we assume the corona to be a point-like, isotropic source located on the
black hole rotation axis, demonstrating the necessity of 
more self-consistent
calculations. We also inspect the effect of motion and geometry of the corona on the emergent spectrum.
Finally, we discuss the implication of anisotropic corona emission for the reflection spectrum in AGNs as well as black hole 
X-ray binaries (BHXRBs). We find that by assuming the corona emission to be isotropic, one may underestimate the soft excess in AGNs
and the reflection continuum and iron K fluorescent line flux in BHXRBs.

\end{abstract}

\keywords{methods: numerical --- radiative transfer --- relativistic processes --- galaxies: active}


\section{Introduction} 
The hard X-ray ($\gtrsim 2~\rm keV$) spectrum of active galactic nuclei (AGNs) 
is usually dominated by a non-thermal component with a high-energy
cut-off at tens to hundreds of keV \citep[e.g.,][]{fabian_properties_2017,tortosa_nustar_2018}.
The hard X-ray is generally believed to be dominated by radiation from the corona, which contains hot
(with a temperature of tens to hundreds of keV), 
optically-thin plasma that Comptonises UV/optical disc photons \citep[e.g.,][]{haardt_two-phase_1991}.
The corona emission also irradiates the cold disc to produce the reflection emission \citep[see][and references therein]{fabian_x-ray_2010}.

For optically-thin thermal plasma Comptonising low-frequency thermal radiation, above $\sim3 k_{\rm B}T_{\rm bb}$ (where $k_{\rm B}$ is
the Boltzmann constant and $T_{\rm bb}$ is the temperature of the thermal radiation) the spectrum is expected to be a powerlaw with
a high-energy cut-off \citep[e.g.][]{sunyaev_comptonization_1980}.
As shown by analytical studies and Monte Carlo simulations,
the slope of the powerlaw is a function of the electron temperature and the optical depth 
\citep[e.g.,][]{sunyaev_comptonization_1980,pozdnyakov_comptonization_1983}
while the cut-off energy depends sensitively on the electron temperature.
Hence the electron temperature and optical depth of the corona can be constrained by analysing the hard X-ray spectrum.
In contrast, the size and geometry of the corona are still not clear. 

A constraint on the size of the corona may offer us the opportunity to discriminate models for formation of AGN corona,
as different mechanisms predict a distinct size of the corona.
For example, we would expect a compact corona if the formation of the corona is due to the ``aborted'' jet \citep[e.g.,][]{ghisellini_aborted_2004}.
In contrast in the ``two-phase'' corona model an extended corona is expected \citep{haardt_two-phase_1991}.
A constraint on the size of the corona will also tell us if some physical processes, such as pair production,
are important \citep[e.g.,][]{guilbert_spectral_1983,zdziarski_nonthermal_1985,fabian_properties_2015}.

To date, the most promising constraint on the size of the corona in AGNs comes from analysis of strongly lensed quasars,
where the microlensing by stellar components in the lensing galaxy results in a complex
magnification pattern \citep[see the review of][]{wambsganss_part_2006}. The relative motion of the lensed quasar,
the galaxy and its stellar components, and the observer
leads to uncorrelated variability. The variability amplitude depends sensitively on the size of the emitting region, with larger
amplitude from smaller emitting region. Thus the size of the source region can be estimated by modelling the light curve
\citep[e.g.][]{kochanek_quantitative_2004,kochanek_turning_2007}. 
\citet{reis_size_2013} compiled measurements of the X-ray emitting region in lensed quasars
and found the sizes to be in the range of $\sim1$ to tens of gravitational radii.
However, the paucity of strongly lensed quasars limits the application of this method.

\citet[hereafter DD16]{dovciak_minimum_2016} proposed a method to estimate the size of the corona in AGNs with simultaneous UV/X-ray observations.
In this method one first obtains the mass accretion rate with UV luminosity, and then calculates the seed photon spectrum received by
a lamp-post corona at an assumed height. They found out that the spectrum is close to blackbody, such that 
\textsc{nthcomp}\footnote{https://heasarc.gsfc.nasa.gov/xanadu/xspec/manual/node200.html}\citep{zdziarski_broad-band_1996,zycki_1989_1999} 
can be utilised to calculate the Comptonised spectrum.
Given that Comptonisation conserves the number of photons (neglecting double Compton effect), one can calculate
the X-ray flux density. Together with the observed X-ray luminosity, a constraint can be put on the size of the corona.


The \citetalias{dovciak_minimum_2016} method uses \textsc{nthcomp}, which assumes that the seed photons are
illuminating the corona isotropically. Apparently this is not the case for AGN corona and as we will show later, 
anisotropic illumination of the seed photons would lead to anisotropic emission of AGN corona.
Another limitation in the \citetalias{dovciak_minimum_2016} method is that the calculation of the Comptonised flux is made 
assuming a lamp-post corona, thus
implicitly assuming that the spectra do not change much across the extended corona. This assumption may not hold for a large corona.
To perform a more realistic, self-consistent modelling of corona emission, we develop 
\codename{}, a general relativistic Monte Carlo code dedicated to calculations of Comptonised spectra in the
Kerr spacetime.
We assume Klein-Nishina cross section and take all general relativistic effects into account.


In general the polarisation of emission is sensitive to the geometry of the source. For AGN corona,
X-ray polarimetry may play an important role
in breaking the degeneracy of corona shape \citep[e.g.,][]{schnittman_x-ray_2010,beheshtipour_x-ray_2017,tamborra_moca_2018}. 
In \codename{} we also have the option of performing polarised radiative transfer. If this option is switched on, we also
propagate the polarisation vector along the null geodesic
and take into account the change of photon polarisation angle and degree due to scattering.

\section{Procedure}
We follow \citet{dolence_grmonty_2009} to account for disc emissivity with the ``superphotons'' scheme.
One ``superphoton'' is a package of several identical photons. Each
superphoton is parameterised by: its energy at infinity $E_\infty$, weight $w$, and polarisation degree $\delta$, which are
constants of motion; four-position $x^\mu$, wave vector $k^\mu$, and polarisation vector $f^\mu$, which need to be propagated along the geodesic.
The weight $w$ has the physical meaning of photon generation rate per unit time in a distant observer's frame.

We first generate superphotons according to disc emissivity, and then ray-trace the superphotons along null geodesics.
While the superphoton is inside the corona, we set the step size of ray-tracing to be much less than the scattering mean free path, and for
each step we evaluate the scattering optical depth covariantly.
If the superphoton is scattered, we sample the differential cross section and calculate the energy, momentum,
and polarisation properties after scattering. 
The propagation terminates if the superphoton enters the event horizon, arrives at infinity, or hits the disc.

The energy and polarisation spectra can be reconstructed by counting superphotons arriving at infinity.
For an observer at inclination $i$, the luminosity density (in counts per unit time per energy interval) at energy $E$ is
\begin{equation}
 L_E = \frac{4\pi\sum_k w_k}{\Delta E \Delta \Omega},
\end{equation}
where the sum is performed over all superphotons with $\theta_{\infty} \in (i - \Delta i/2, i + \Delta i /2]$ and $E_{\infty} \in (E - \Delta E / 2,
E + \Delta E / 2]$, $\Delta E$ is the width of the energy bin, $\Delta \Omega = 2\pi [{\rm cos}(i - \Delta i /2) - {\rm cos}(i + \Delta i /2)]$ is
the solid angle extended by the inclination bin, and $E_\infty$, $\theta_\infty$ are the energy and polar angle of the superphotons at infinity, respectively.
As Compton scattering only induces linear polarisation, we always let the Stokes parameter $V=0$. For the other two Stokes parameters:
\begin{eqnarray}
Q_E = \frac{4\pi\sum_k w_k\delta_k\ {\rm cos}2\psi_k}{\Delta E \Delta \Omega}, \\
U_E = \frac{4\pi\sum_k w_k\delta_k\ {\rm sin}2\psi_k}{\Delta E \Delta \Omega},
\end{eqnarray}
where $\psi$ is the polarisation angle measured at infinity in a comoving frame attached to the observer, while $x-$ and $y-$axes are identified as
$\partial/\partial\theta$ and $\partial/\partial\phi$, respectively.

\iftrue
\subsection{The Kerr metric}
The covariant Kerr metric in the Boyer-Lindquist coordinates:
\begin{eqnarray}
g_{\mu\nu}  & = &
 \left[
\begin{array}{cccc}
-\left(1 - \frac{2r}{\rho^2}\right) & 0 & 0 & -\frac{2ar{\rm sin}^2\theta}{\rho^2}\\
0 & \frac{\rho^2}{\Delta} & 0 & 0\\
0 & 0 & \rho^2 & 0 \\
-\frac{2ar{\rm sin}^2\theta}{\rho^2} & 0 & 0 & \frac{A{\rm sin}^2\theta}{\rho^2}
\end{array}     \right] ,
\end{eqnarray}
where $\rho^2 \equiv r^2 + a^2 {\rm cos}^2\theta$, $\Delta \equiv r^2 - 2r + a^2$,
and $A \equiv (r^2 + a^2)^2 - \Delta a^2 {\rm sin}^2 \theta$. Here we let $G=M=c=1$.

As the Kerr metric is stationary and axially symmetric, we immediately obtain two constants of motion:
the photon energy at infinity $E_\infty \equiv -p_t$, and the photon angular momentum about the symmetric axis $L_z \equiv p_\phi$. 
\citet{carter_global_1968} found another constant of motion $\mathcal{Q} \equiv p_\theta^2 - a^2 {\rm cos}^2\theta + L_z^2 {\rm cot}^2\theta$. We denote 
$l\equiv L_z / E_\infty$, $Q\equiv \mathcal{Q} / E^2_\infty$.
\fi

\subsection{Ray-tracing in the Kerr spacetime}
In \codename{} we implement two independent methods of ray-tracing photons along null geodesics:
the first one is to integrate the geodesic equation directly,
while the second one is to make use of the separability of the Hamilton-Jacobi equation and integrate the equations of motion of $r$ and $\theta$.
We find these two methods to give consistent results.

\subsubsection{Integration of the geodesic equation}
The propagation of photon is described by:
\begin{equation}
\frac{dx^\mu}{d\lambda} = k^\mu,
\end{equation}
where $\lambda$ is the affine parameter.
We normalise $\lambda$ in such a way that $k_t = -1$. The propagation of $k^\mu$ is governed by the geodesic equation:
\begin{equation}
 \frac{dk^\mu}{d\lambda} = -\Gamma^\mu_{\nu\sigma} k^\nu k^\sigma,
\end{equation}
where $\Gamma^\mu_{\nu\sigma}$ are the Christoffel symbols.
We utilise the \textsc{sim5} package \citep{bursa_numerical_2017} to perform the ray-tracing,
which uses the Verlet algorithm to 
evaluate the integral, following \citet{dolence_grmonty_2009}.

\subsubsection{Equations of motion after separation of the Hamilton-Jacobi equation}
The Hamilton-Jacobi equation of the Kerr spacetime is separable \citep{carter_global_1968}. As a result, we can relate the following
two integrals of $r$ and $\mu \equiv {\rm cos}\theta$, respectively, where $r$ and $\theta$ are two Boyer-Lindquist coordinates
\citep{chandrasekhar_mathematical_1983}:
\begin{equation}
\label{eq:rmu}
 I = s_r \int \frac{dr}{\sqrt{R(r)}} = s_\mu\int \frac{d\mu}{\sqrt{M(\mu)}}, 
\end{equation}
where $s_r,s_\mu=\pm1$. We set $s_r$ and $s_\mu$ in such a way that the integrals are always positive.
For massless particles
\begin{eqnarray}
 R(r) &=& r^4 + (a^2 - l^2 - Q) r^2 + 2\left[Q + (l - a)^2\right] r \nonumber \\
  &&- a^2 Q,\\
 M(\mu) &=& Q + (a^2 - l^2 - Q)\mu^2  - a^2 \mu^4.
\end{eqnarray}

Starting with a superphoton at $x^{\mu}_0$ with wave vector $k^\mu_0$ and setting $I=0$, we can ray-trace the photon trajectory by the following way: 
for each step we give $I$ a small
increment such that $I\rightarrow I + dI$, and then invert the integrations in Eq.~\ref{eq:rmu} to solve for $r$ and $\mu$. We also 
change signs of $k^r$ and $k^\theta$ upon encountering the respective turning points. At any step
one can solve for $k^\mu$ given $l$, $Q$, and signs of $k^r$ and $k^{\theta}$ following \citet{carter_global_1968}.
The affine parameter can be obtained by evaluating the following equation \citep{carter_global_1968}:
\begin{equation}
\label{eq:lambda}
 \lambda = \int \frac{r^2dr}{\sqrt{R(r)}} + a^2 \int \frac{\mu^2 d\mu}{\sqrt{M(\mu)}}.
\end{equation}
In this paper we are dealing with axial-symmetric problems and we do not study variability properties, hence evaluation of $t$ and $\phi$ is not included.
Integrals in Eqs.~\ref{eq:rmu} and \ref{eq:lambda} are evaluated by making use of Carlson's elliptical functions, following
\citet{rauch_optical_1994,li_multitemperature_2005,dexter_fast_2009}.

\subsection{Propagation of the polarisation vector in the Kerr spacetime}
It is known that Compton scattering only induces linear polarisation.
For linear polarisation, the polarisation state can be
described by the polarisation degree $\delta$ that is an invariant scalar, and a real vector 
$f^\mu \equiv A^\mu / A$ where $A^\mu$ is the wave amplitude vector \citep{misner_gravitation_1973}. The polarisation vector $f^\mu$ 
is orthonormal to the photon wave vector $k^\mu$:
\begin{eqnarray}
\label{eq:polvec_orth}
 f^\mu f_\mu = 1,\\
 k^\mu f_\mu = 0,
\end{eqnarray}
and is parallel transported along the null geodesic:
\begin{equation}
\label{eq:polvec_para}
 k^\mu \nabla_\mu f^\nu = 0.
\end{equation}
For such kind of vector, one can find a complex constant: the Walker-Penrose constant $\kappa_{\rm wp}$
that is conserved along the null geodesic \citep{walker_quadratic_1970,chandrasekhar_mathematical_1983}. So instead of propagating $f^\mu$ using
Eq.~\ref{eq:polvec_para}, we just keep a record of $\kappa_{\rm wp}$. Only when the photon interacts with an electron, we solve for $f^\mu$, calculate
$f^\mu$ and $k^{\mu}$ after the interaction, and subsequently re-evaluate $\kappa_{\rm wp}$.

To evaluate $\kappa_{\rm wp}$ given $f^\mu$ and $k^\mu$ \citep{connors_polarization_1980}:
\begin{equation}
\label{eq:kwp}
\begin{split}
 \kappa_{\rm wp} = (r-ia\ {\rm cos}\theta)\{(k^t f^r - k^r f^t) + a\ {\rm sin}^2\theta(k^rf^\phi - k^\phi f^r) \\
 - i[(r^2+a^2)(k^\phi f^\theta - k^\theta f^\phi) - a(k^t f^\theta - k^\theta f^t)]{\rm sin}\theta\}.
\end{split}
\end{equation}

With Eqs.~\ref{eq:polvec_orth}--\ref{eq:polvec_para}, given $\kappa_{\rm wp}$ we can only determine $f^\mu$ up to 
an addition of multiple of $k^\mu$, as these
equations are gauge invariant under the transformation:
\begin{equation}
\label{eq:gauge}
 f^\mu \rightarrow f^\mu + \alpha k^\mu,
\end{equation}
where $\alpha$ is an arbitrary real number. We put an additional constraint that $f^t=0$ and solve for the other
three vector components of $f^\mu$. 

Finally, at infinity we follow \citet{li_inferring_2009} to evaluate the Stokes parameters given $\delta$ and $\kappa_{\rm wp}$.

\subsection{Orthonormal tetrad}
One can construct orthonormal tetrad $\boldsymbol{e}_{(a)}$ for an observer that is comoving with the fluid.
To transform a vector $\boldsymbol{f}$ from the tetrad frame to the coordinate frame,
\begin{equation}
 f^\mu = e^\mu_{(a)} f^{(a)},
\end{equation}
and to transform back,
\begin{equation}
 f^{(a)} = e_\mu^{(a)} f^\mu.
\end{equation}
We illustrate the method to obtain the orthonormal tetrad in the Appendix.

\subsection{Superphoton generation}
To find out the generation rate of superphotons emitted by an optically-thick accretion disc on
the equatorial plane, we first notice that the 
flux density defined as follows is a relativistic invariant \citep{kulkarni_measuring_2011,schnittman_monte_2013}:
\begin{equation}
 F_\nu = I_{\nu} \frac{1}{u^t} {\rm cos}\theta_{\rm em} d\mathcal{S} d\Omega,
\end{equation}
where $I_\nu$ is the specific intensity, 
$\frac{1}{u^t}$ is the factor that transforms time from the disc fluid rest frame into the distant observer's frame,
$\theta_{\rm em}$ is the polar emission angle with respect to disc norm in the disc fluid rest frame,
$d\mathcal{S}$ is the proper area, and $d\Omega$ is the solid angle.
It is easy to see that the number of photons emitted per unit time in the distant observer's frame
\begin{equation}
 \dot{N} = \int \frac{F_\nu}{h\nu} d\nu
\end{equation}
is also an invariant. For photons emitted from an annulus on the equatorial plane that centers at Boyer-Lindquist radius $r$ and has
a width of $dr$:
\begin{equation}
 \frac{1}{u^t} = \sqrt{1 - 2/r + 4\Omega a / r - \Omega^2 (r^2 + a^2 + 2a^2/r)},
\end{equation}
where $\Omega$ is the angular velocity of the disc fluid; and the proper area \citep{wilkins_understanding_2012}:
\begin{equation}
 d\mathcal{S} = \frac{2\pi\rho \Gamma}{\sqrt{\Delta}} \sqrt{r^2 + a^2 + \frac{2a^2 r}{\rho^2}} dr,
\end{equation}
where $\Gamma$ is the Lorentzian factor of disc fluid as measured by a zero angular momentum observer (ZAMO).
For Keplerian discs, 
\begin{equation}
 \Omega = \frac{1}{r^{3/2} + a}.
\end{equation}
The velocity along $\phi$ direction as measured by a stationary observer
\begin{equation}
  v_{\phi} = \frac{(\Omega - \omega)\mathcal{A}}{r^2\sqrt{\Delta}},
\end{equation}
where $\omega$ is the frame-dragging angular velocity;
hence the Lorentzian factor
\begin{equation}
 \Gamma = \frac{1}{\sqrt{1 - v_\phi^2}}.
\end{equation}

If the local spectrum is color-corrected blackbody with a color correction factor of $f_{\rm col}$ and
an effective temperature of $T_{\rm eff}$:
\begin{equation}
I_{\nu} = f_{\rm limb}\frac{1}{f_{\rm col}^4} \frac{2h\nu^3}{c^2}\frac{1}{e^{h\nu/k_{\rm B}f_{\rm col}T_{\rm eff}} - 1},
\end{equation}
where $f_{\rm limb}$ is the limb-darkening factor; we have
\begin{equation}
  \dot{N} = \frac{4\zeta(3) k_{\rm B}^3}{c^2h^3} \frac{f_{\rm limb} {\rm cos}\theta_{\rm em} d\mathcal{S} d\Omega
 T_{\rm eff}^3}{f_{\rm col}u^t},
\end{equation}
where $\zeta$ is the Riemann zeta function.
Therefore from the annulus, we sample $N_s$ photons each with weight $w=\dot{N}/N_s$ into the solid angle $d\Omega$. 
The energies of the sampled photons at infinity follow the Planckian distribution with a temperature of
$g f_{\rm col} T_{\rm eff}$, where $g = -1/k_\mu U^\mu$ is the redshift factor, 
and $U^\mu$ is the disc particle four-velocity.
We use a rejection method as illustrated in details in Section 9.4 of \citet{pozdnyakov_comptonization_1983} to sample the Planckian distribution.

We assume Novikov-Thorne emissivity profile with zero torque at the inner edge of the disc. The local spectrum is color corrected blackbody.
If the polarisation option is switched off, we assume that the disc emission is isotropic in the local frame, and the seed superphotons emitted by the
disc are unpolarised.
Otherwise we model the disc atmosphere as a semi-infinite plane atmosphere \citep{chandrasekhar_radiative_1960},
as in \citet{dovciak_thermal_2008,li_inferring_2009}, to calculate the polarisation properties and the
angular distribution of the seed superphotons.
In this case the polarisation angle is perpendicular to the meridian plane, the plane 
contains both the symmetry axis and the line of sight to the observer; and the polarisation degree increases monotonically with the polar emission angle.
Given the emission angle, 
the intensity and polarisation degree of the radiation can be calculated using Chandrasekhar's H-functions \citep{chandrasekhar_radiative_1960}.
We utilise the method described in \citet{bosma_efficient_1983} to evaluate the H-functions.


\subsection{Comptonisation}
While the superphoton is travelling inside the corona, for each step we evaluate the scattering probability
assuming Klein-Nishina cross section. If the superphoton is scattered, 
first we sample the momentum of the scattering electron and boost $k^\mu$ and $f^\mu$ into the electron rest frame.
Then we sample the momentum of the scattered photon and calculate the polarisation vector after scattering. Finally we boost $k^\mu$ and $f^\mu$
back into the Boyer-Lindquist frame and update $E$, $\delta$, and $k_{\rm wp}$ of the superphoton. If the polarisation option is switched off, 
we always assume that the incoming superphoton is unpolarised.

\subsubsection{Covariant evaluation of the scattering optical depth}
While we propagate the superphoton inside the corona, for each step we evaluate the covariant scattering optical depth following \citet{younsi_general_2012}:
\begin{equation}
 \tau_\nu = -\int_{\lambda_0}^{\lambda_1} \alpha_{0,\nu}(\lambda) k_\alpha U^\alpha|_\lambda d\lambda,
\end{equation}
where $\lambda_0$ and $\lambda_1$ are the affine parameters of the superphoton at the beginning and the end of the step,
and $\alpha_{0,\nu}$ is the scattering coefficient in the fluid rest frame.

The superphoton energy measured in the fluid frame is $E_0 = -k_\mu U^\mu E_\infty$. Denoting $x_0\equiv E_0/(m_e c^2)$ 
where $m_e$ is the electron rest mass, 
the scattering coefficient of the superphoton
with respective to a population of electron can be evaluated via the following integral \citep[e.g.,][]{pozdnyakov_comptonization_1983}:
\begin{equation}
\label{eq:alpha0}
 \alpha_{0,\nu} =\int \frac{dN_e}{d^3p} (1-\mu_e \beta_e) \sigma(x) d^3p,
\end{equation}
where $\frac{dN_e}{d^3p}$ is the electron velocity distribution, 
$\mu_e ={\rm cos}\theta_e$ while $\theta_e$ is the angle between the momenta of the photon and
the electron, $\beta_e \equiv v_e/c$ is the electron velocity, $\gamma_e$ is the Lorentz factor of the electron,
$\sigma(x)$ is the scattering cross section, and $x=\gamma_e x_0(1-\mu_e \beta_e)$ is the dimensionless photon energy
in the electron rest frame.

In the electron rest frame, the differential scattering cross section is \citep{berestetskii_relativistic_1971,connors_polarization_1980}:
\begin{equation}
\label{eq:kndiff_pol}
\begin{aligned}
 \frac{d\sigma}{d\Omega} = & \frac{r_0^2}{2}\left(\frac{x^\prime}{x}\right)^2
 \biggl[\frac{x}{x^\prime} + \frac{x^\prime}{x}  - {\rm sin}^2 \theta^\prime \\
  & - {\rm sin}^2\theta^\prime(X_s {\rm cos}2\phi^\prime + 
 Y_s {\rm sin}2\phi^\prime)\biggr],
\end{aligned}
\end{equation}
where $r_0$ is the classical electron radius, $x^\prime$ is photon energy after scattering,
and $\theta^\prime$ and $\phi^\prime$ are the polar and azimuthal angles of the
photon wave vector after scattering. The coordinate system is defined in such a way that
the $z-$axis is aligned with the photon wave vector, while the $x-$ and $
y-$axes are defined by two orthonormal unit vectors in the plane perpendicular to the photon wave vector.
For a light beam with polarisation degree $\delta$ and
polarisation angle $\psi$:
\begin{equation}
 X_s = \delta {\rm cos}2\psi, \ Y_s = \delta{\rm sin}2\psi.
\end{equation}
By integrating Eq.~\ref{eq:kndiff_pol}, we obtain the total cross section:
\begin{equation}
\label{eq:kn}
\begin{aligned}
 \sigma(x) = & \pi r_0^2 \frac{1}{x}\biggl[\left(1 - \frac{2}{x} - \frac{2}{x^2}\right){\rm ln} (1 + 2x) + \frac{1}{2} + \frac{4}{x} \\
    & - \frac{1}{2(1+2x)^2}\biggr],
 \end{aligned}
\end{equation}
which is the Klein-Nishina formula $\sigma_{\rm KN}$ and is independent of the polarisation degree.

For thermal electrons with temperature $T$, their velocities follow the Maxwell-J\"uttner distribution:
\begin{equation}
\label{eq:maxwell}
 \frac{dN_e}{d\gamma_e} \propto \frac{\gamma_e^2\beta_e}{\theta_T K_2(1/\theta_T)}e^{-\gamma_e/\theta_T},
\end{equation}
where $\theta_T\equiv k_{\rm B}T/m_e c^2$ is the dimensionless electron temperature, and $K_2$ is the modified Bessel function of order 2.

With Eqs.~\ref{eq:alpha0}, \ref{eq:kn}, and \ref{eq:maxwell} we can calculate the optical depth $\tau$ in the case of thermal electron. 
With optical depth $\tau$, the scattering probability is $P=1-e^{-\tau}$. For optically-thin coronae,
we introduce a bias factor $b \gg 1$ following the
practice of \citet{pozdnyakov_comptonization_1983,dolence_grmonty_2009} to enhance the statistics at high energy,
such that the scattering probability becomes $P=1-e^{-b\tau}$, and after scattering
the superphoton ``splits'' into two superphotons with appropriate weights. 
We generate a random number $\epsilon$ by sampling a uniform distribution between 0 and 1, and the condition for scattering is $\epsilon \leq P$.



\subsubsection{Sampling electron momentum}
If the photon is scattered, we first sample the momentum of the scattering electron.
The probability density of scattering electron to have momentum $\boldsymbol{p}_e = [\gamma_e m_e, \gamma_e m_e \boldsymbol{v}]$:
\begin{equation}
 P(\boldsymbol{p}_e)\propto \frac{dN_e}{d^3p} (1-\mu_e \beta_e)\sigma_{\rm KN}.
\end{equation}
We sample the probability density distribution with a rejection method, following \citet{pozdnyakov_comptonization_1983,canfield_inverse_1987}. 

\subsection{Scattering in the electron rest frame}
Once we sample the four-velocity of the scattering electron, we can obtain the energy, the wave vector, and the 
polarisation vector of the photon in the
electron rest frame with a generic Lorentz transformation. 
For the polarisation vector in the electron rest frame $\boldsymbol{f}_e$ we make an additional transformation according to Eq.~\ref{eq:gauge},
in which we set $\alpha$ in such a way that $f_e^t=0$, 
as required by the gauge we choose. 

For convenience we set up such a 3D Cartesian coordinate system in which the $z-$ and $x-$ axes are aligned with the wave and polarisation vectors of
the incoming photons, respectively. 
The differential cross section for the photon to be scattered into a solid angle $d\Omega^\prime$ centering at polar angle $\theta^\prime$
and azimuthal angle $\phi^\prime$ is
\begin{equation}
 \frac{d\sigma}{d\Omega^\prime} = \frac{r_0^2}{2}\left(\frac{x^\prime}{x}\right)^2
 \left[\frac{x}{x^\prime} + \frac{x^\prime}{x} - {\rm sin}^2 \theta^\prime - \delta{\rm sin}^2\theta^\prime {\rm cos}2\phi^\prime\right],
\end{equation}
where $x^\prime$, the dimensionless photon energy after scattering, is related with $\theta^\prime$ by the Compton recoil relation:
\begin{equation}
x^\prime = \frac{x}{1 + x(1-{\rm cos}\theta^\prime)}.
\end{equation}
Hence the joint probability density of $x^\prime$ and $\phi^\prime$ is:
\begin{equation}
\label{eq:kndiff_pol1}
 p(x^\prime,\phi^\prime;\delta) = \frac{1}{\sigma_{\rm KN}} \frac{d\sigma}{d\Omega^\prime}\frac{d{\rm cos}\theta^\prime}{dx^\prime}.
\end{equation}
By integrating Eq.~\ref{eq:kndiff_pol1} over $\phi^\prime$,
we obtain the probability density of $x^\prime$:
\begin{equation}
 p(x^\prime) = \frac{1}{\sigma_{\rm KN}} \frac{\pi r_0^2}{x^2}
 \left[\frac{x}{x^\prime} + \frac{x^\prime}{x} - {\rm sin}^2 \theta^\prime\right],
\end{equation}
which is independent of $\delta$. We sample $x^\prime$ following \citet{kahn_applications_1954}, and subsequently solve for the 
scattering polar angle $\theta^\prime$ with the Compton recoil relation.
The probability density distribution of the scattering azimuthal angle $\phi^\prime$:
\begin{equation}
 p(\phi^\prime | x^\prime ; \delta) = \frac{p({x^\prime, \phi^\prime; \delta})}{p({x^\prime})} 
 = \frac{1}{2\pi} - \frac{\delta {\rm sin}^2\theta^\prime {\rm cos}2\phi^\prime}
 {2\pi \left(\frac{x}{x^\prime} + \frac{x^\prime}{x} - {\rm sin}^2\theta^\prime\right)}.
\end{equation}
It is trivial to calculate the cumulative distribution of $\phi^\prime$:
\begin{equation}
{\rm cdf}(\phi^\prime) = \frac{\phi^\prime}{2\pi} - \frac{\delta {\rm sin}^2\theta^\prime {\rm sin}2\phi^\prime}
 {4\pi \left(\frac{x}{x^\prime} + \frac{x^\prime}{x} - {\rm sin}^2\theta^\prime\right)}.
\end{equation}

With $\theta^\prime$ and $\phi^\prime$ we can derive the photon wave vector in the electron rest frame after scattering $\boldsymbol{k}^\prime_e$.
We set up a new coordinate system, with the z-axis aligned with $\boldsymbol{k}^\prime_e$ and
\begin{eqnarray}
\boldsymbol{e}^\prime_{\perp}&=& \frac{\boldsymbol{k}_e \times \boldsymbol{k}^\prime_e}{\sqrt{|\boldsymbol{k}_e \times \boldsymbol{k}^\prime_e|}},\\
\boldsymbol{e}^\prime_{\parallel} &=& \boldsymbol{k}^\prime_e \times \boldsymbol{e}_{\perp}^\prime,
\end{eqnarray}
such that the $\boldsymbol{e}^\prime_{\parallel}$ is the unit vector in the scattering plane and perpendicular to $\boldsymbol{k}^\prime$
while $\boldsymbol{e}^\prime_{\perp}$ is the unit vector perpendicular to the scattering plane. The two normalised Stokes parameters of the scattered photon
in the coordinate system $\{\boldsymbol{e}_\parallel^\prime, \boldsymbol{e}_\perp^\prime\}$ is \citep{connors_polarization_1980}:
\begin{eqnarray}
\label{eq:knstokes}
 \frac{Q^\prime}{I^\prime} &=& \frac{1}{N} [{\rm sin}^2\theta^\prime - \delta(1 + {\rm cos}^2\theta^\prime) {\rm cos}2\phi^\prime],\\
 \frac{U^\prime}{I^\prime} &=&\frac{2}{N} \delta {\rm cos}\theta^\prime {\rm sin}2\phi^\prime,
\end{eqnarray}
where 
\begin{equation}
 N = \frac{x^\prime}{x} + \frac{x}{x^\prime} -{\rm sin}^2\theta^\prime - \delta {\rm sin}^2\theta^\prime {\rm cos}2\phi^\prime.
\end{equation}
The polarisation degree and angle after scattering:
\begin{eqnarray}
 \delta^\prime &=& \frac{\sqrt{Q^{\prime 2} + U^{\prime 2}}} {I^\prime},\\
 \psi^\prime &=& \frac{1}{2}{\rm arctan}\left(\frac{Q^\prime}{U^\prime}\right).
\end{eqnarray}
The polarisation vector is then 
$\boldsymbol{f}_e^\prime = {\rm cos}\psi^\prime \boldsymbol{e}_\parallel^\prime + {\rm sin}\psi^\prime \boldsymbol{e}_\perp^\prime$.
The energy, wave and polarisation vectors after scattering in the Boyer-Lindquist frame can be obtained from $x^\prime$, $\boldsymbol{k}_e^\prime$, and 
$\boldsymbol{f}_e^\prime$ by a series of transformations including
rotation, generic Lorentz boost, and tetrad transform. 
At last we can evaluate the Walker-Penrose constant of the superphoton after scattering following Eq.~\ref{eq:kwp}.



\section{Comparison with previous codes}
\subsection{Spectrum and polarisation from Novikov-Thorne disc}
\begin{figure}
 \includegraphics[width=\columnwidth]{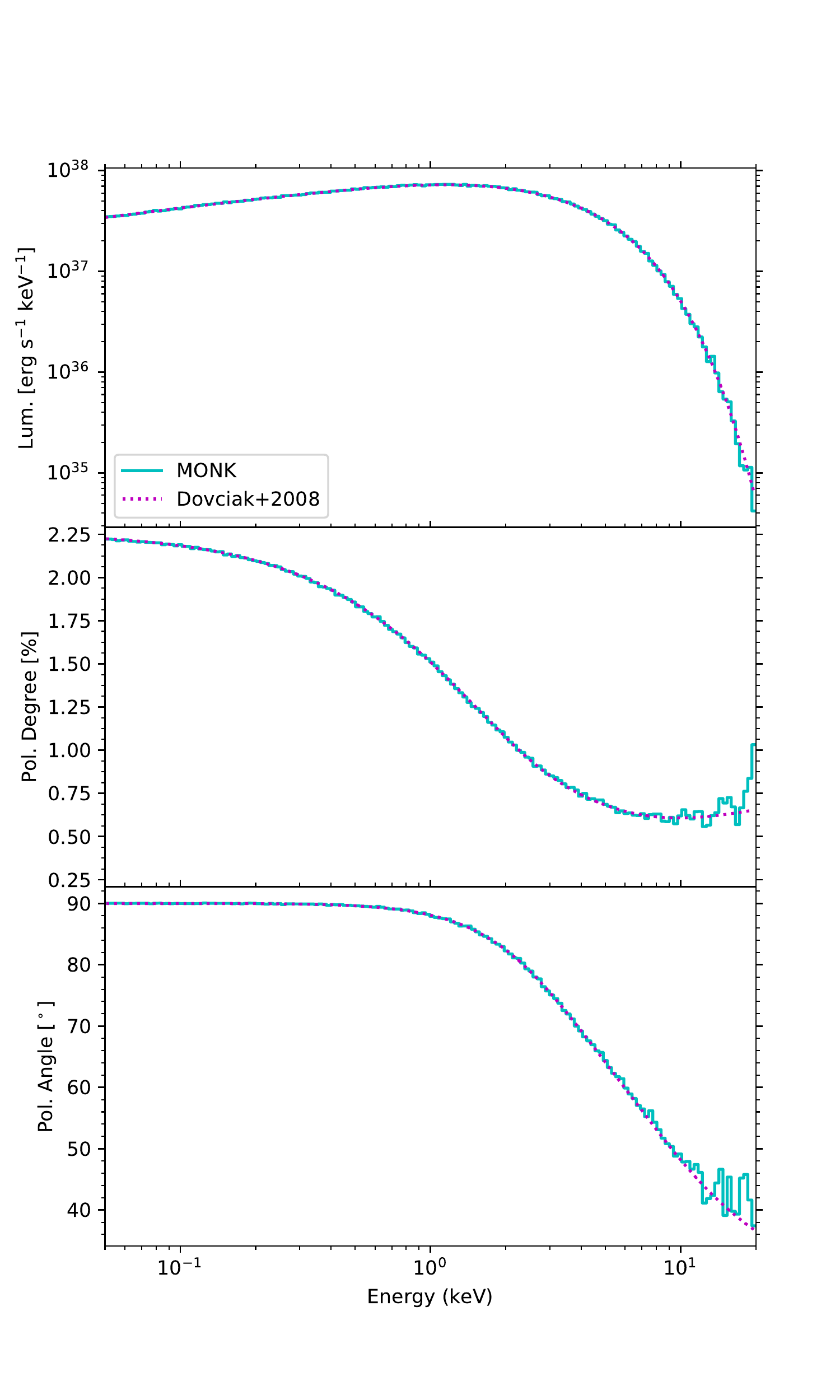}
 \caption{The energy and polarisation spectra from an optically-thin disc around a stellar mass black hole. Parameters:
 $a=1$, $M=14~\rm M_\odot$, $\dot{M}=1.4\times10^{18}~\rm g\ s^{-1}$, and $f_{\rm col}=1.7$. The results from \codename{} and 
 \citet{dovciak_thermal_2008} are plotted in cyan and magenta colors, respectively. From top to bottom: the energy spectrum, 
 the polarisation degree, and the polarisation angle.\label{fig:ntpol}}
\end{figure}

In this section we compare the energy and polarisation spectra of a Novikov-Thorne disc around a stellar-mass black hole with
\citet{dovciak_thermal_2008}. The parameters are: $a=1$, $M=14~\rm M_\odot$, $\dot{M}=1.4\times10^{18}~\rm g\ s^{-1}$, and $f_{\rm col}=1.7$. 
The observer's inclination is $60^\circ$. In Fig.~\ref{fig:ntpol} we present the results
calculated by \codename{} with that by \citet{dovciak_thermal_2008}, and find them to agree well.

\subsection{Energy spectra of spherical plasma cloud Comptonising low-frequency radiation}
\label{sec:srsphere}
In this section we compare the energy spectra of spherical plasma clouds Comptonising low-frequency 
radiation calculated by \codename{} with that calculated by \textsc{grmonty} \citep{dolence_grmonty_2009}.
The isotropic, low-frequency primary radiation is located in the center of the cloud.
Both of the plasma and the primary radiation are thermal, and their temperatures
are $4~\rm m_ec^2$ and $10^{-8}~\rm m_e c^2$, respectively. For comparison we take three different optical depths:
$\tau_{\rm T}=10^{-3},\ 0.1$, and $3$, corresponding to Fig. 7--9 of \citet{dolence_grmonty_2009}. 
Here $\tau_{\rm T}\equiv n_e \sigma_{T} R$, where $n_e$ is the electron number density,
and $R$ is the radius of the cloud.

\begin{figure}
 \begin{center}
 \includegraphics[width=\columnwidth]{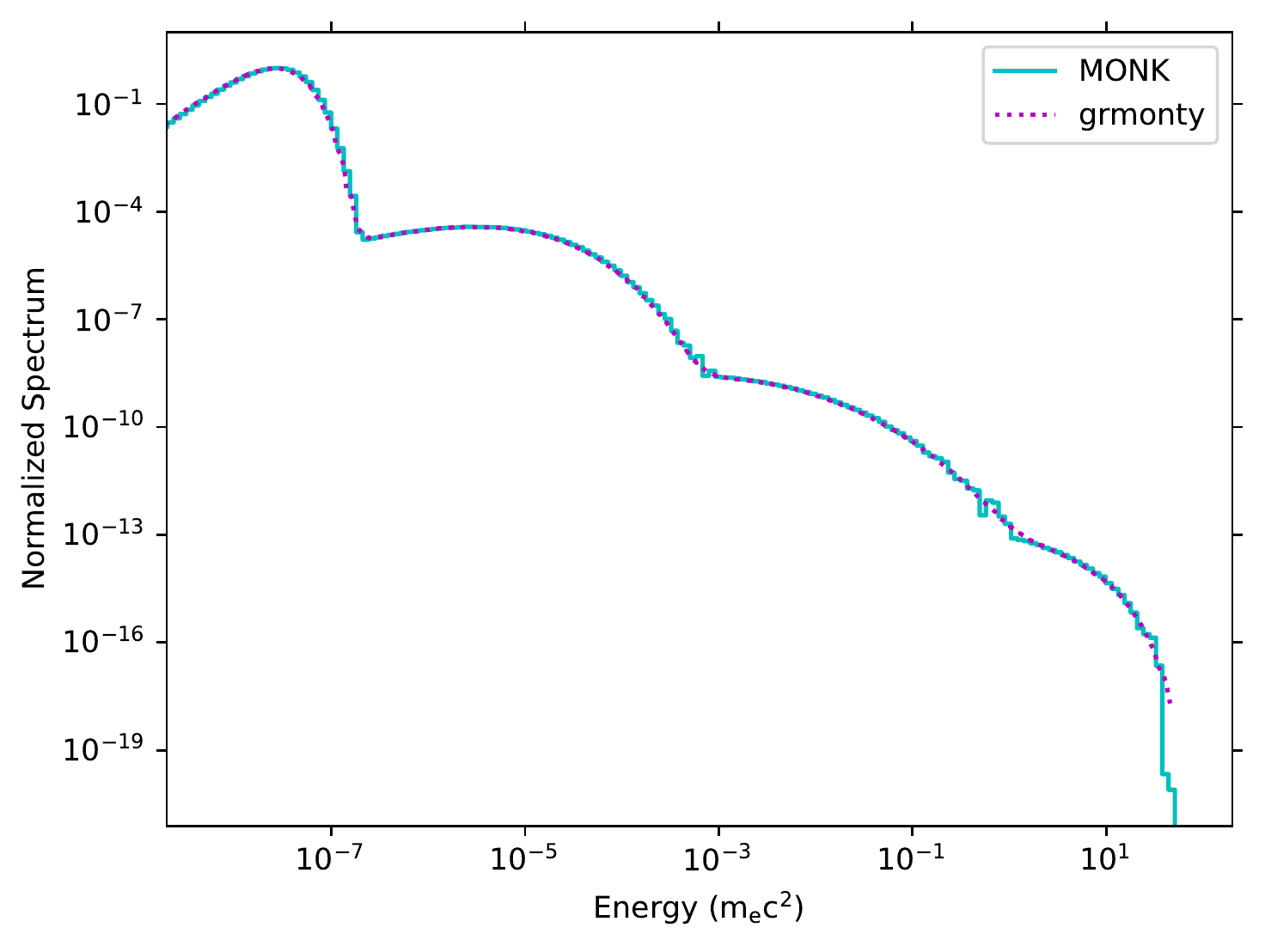}
 \caption{The spectra of a spherical plasma cloud ($T_e=4~\rm m_e~c^2$) Comptonising a central photon source ($T_{\rm bb} = 10^{-8}~\rm m_e~c^2$).
 The Thomson optical optical $\tau_{\rm T} = n_e \sigma_{\rm T} R = 10^{-3}$, where $n_e$ is the electron number density, 
 $\sigma_{T}$ is the Thomson scattering cross section, and $R$ is the radius of the cloud. 
 The cyan solid and magenta dotted lines represent spectra by \codename{} and
 \textsc{grmonty}, respectively.\label{fig:srsphere_1em4}}
 \end{center}
\end{figure}

\begin{figure}
 \begin{center}
 \includegraphics[width=\columnwidth]{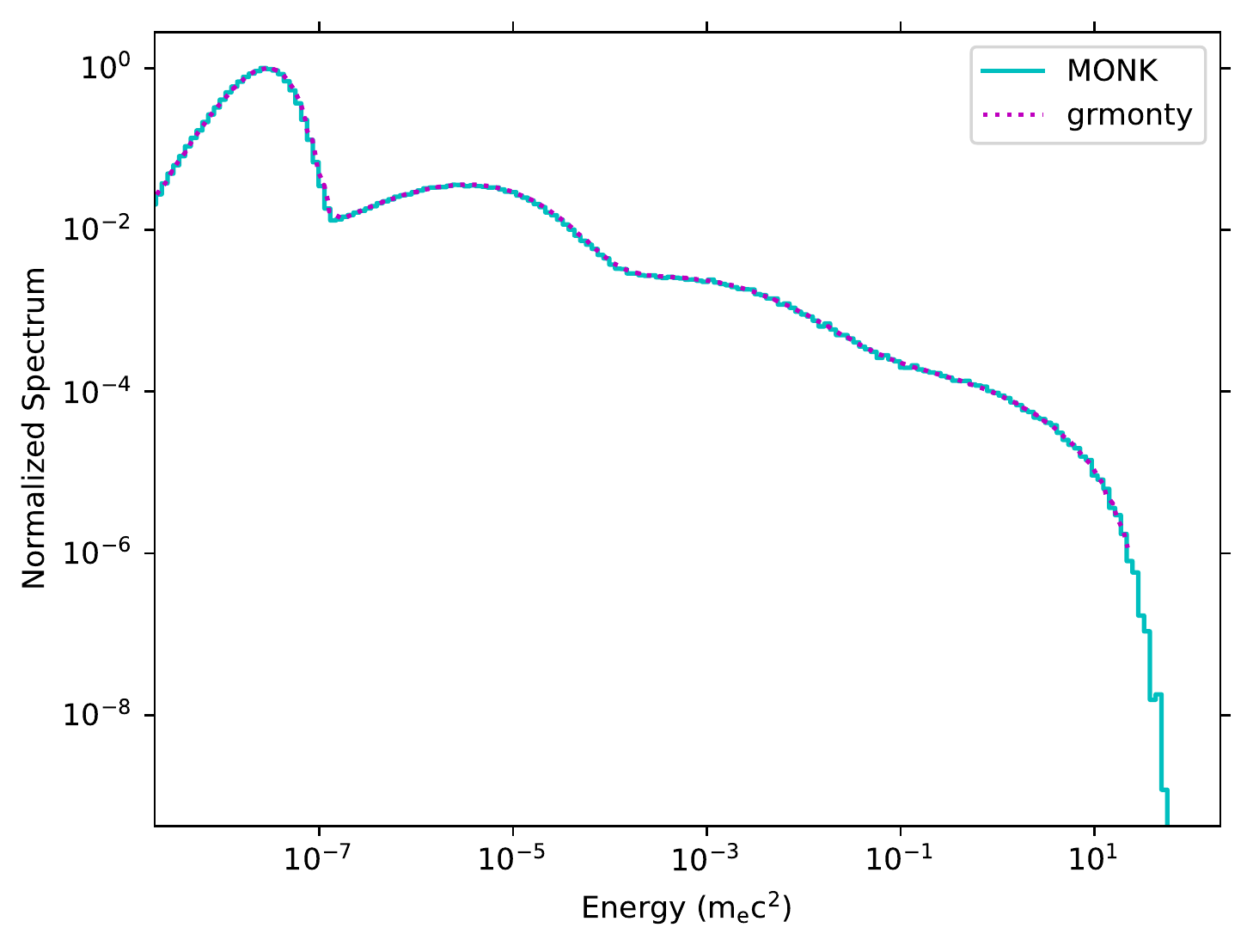}
 \caption{The same with Fig.~\ref{fig:srsphere_1em4}, but for $\tau_{\rm T}=0.1$.\label{fig:srsphere_0p1}}
 \end{center}
\end{figure}

\begin{figure}
 \begin{center}
 \includegraphics[width=\columnwidth]{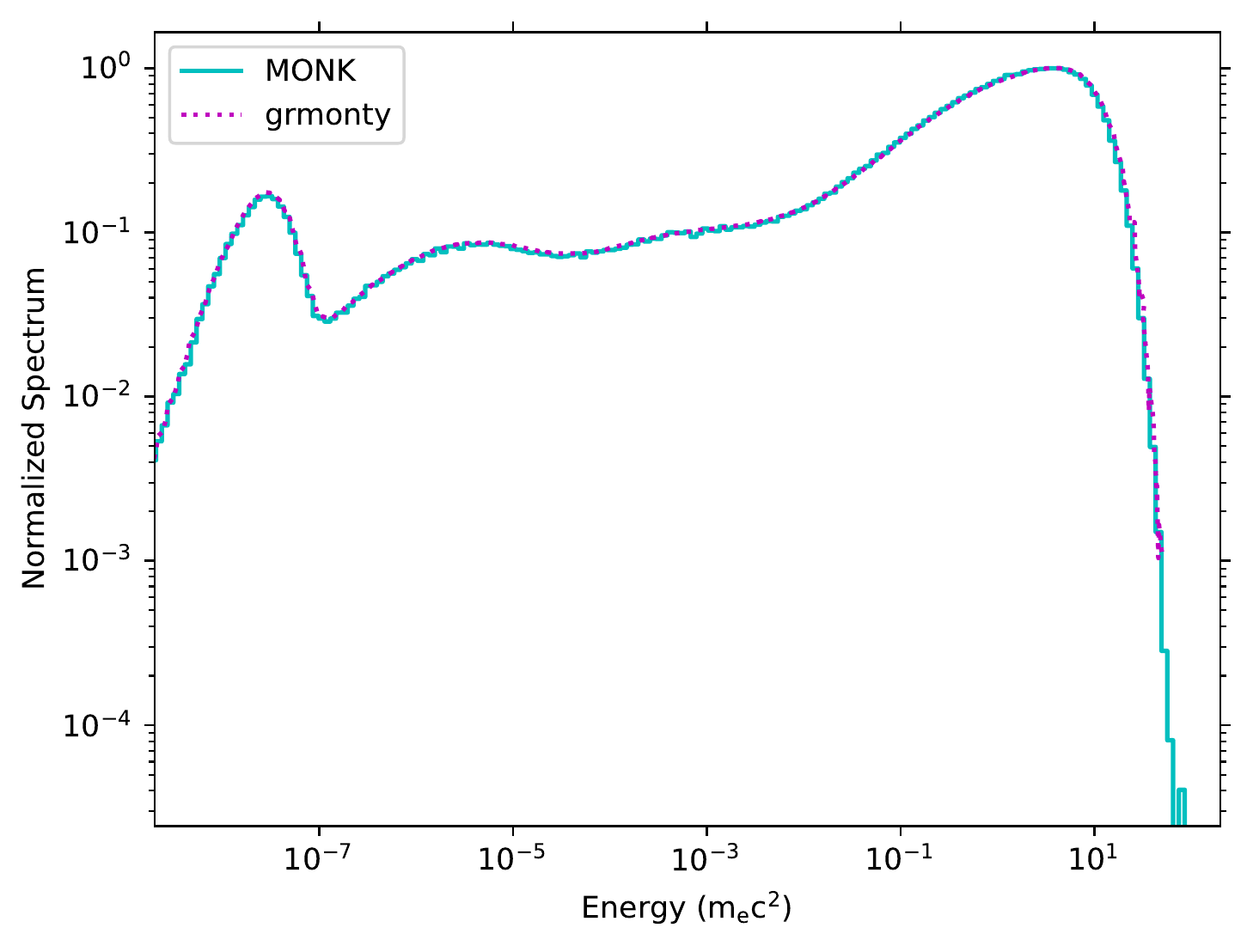}
 \caption{The same with Fig.~\ref{fig:srsphere_1em4}, but for $\tau_{\rm T}=3$.\label{fig:srsphere_3}}
 \end{center}
\end{figure}

In Figs.~\ref{fig:srsphere_1em4}--\ref{fig:srsphere_3} we present the spectra calculated with \codename{} and \textsc{grmonty}, and find
them in good agreement.
The spectra by \textsc{grmonty} are read from their paper using \textsc{WebPlotDigitizer}\footnote{https://automeris.io/WebPlotDigitizer/}.
The polarisation option is turned off.

%

\subsection{Polarization degree of radiation from scattering disc atmosphere}
\begin{figure}
 \includegraphics[width=\columnwidth]{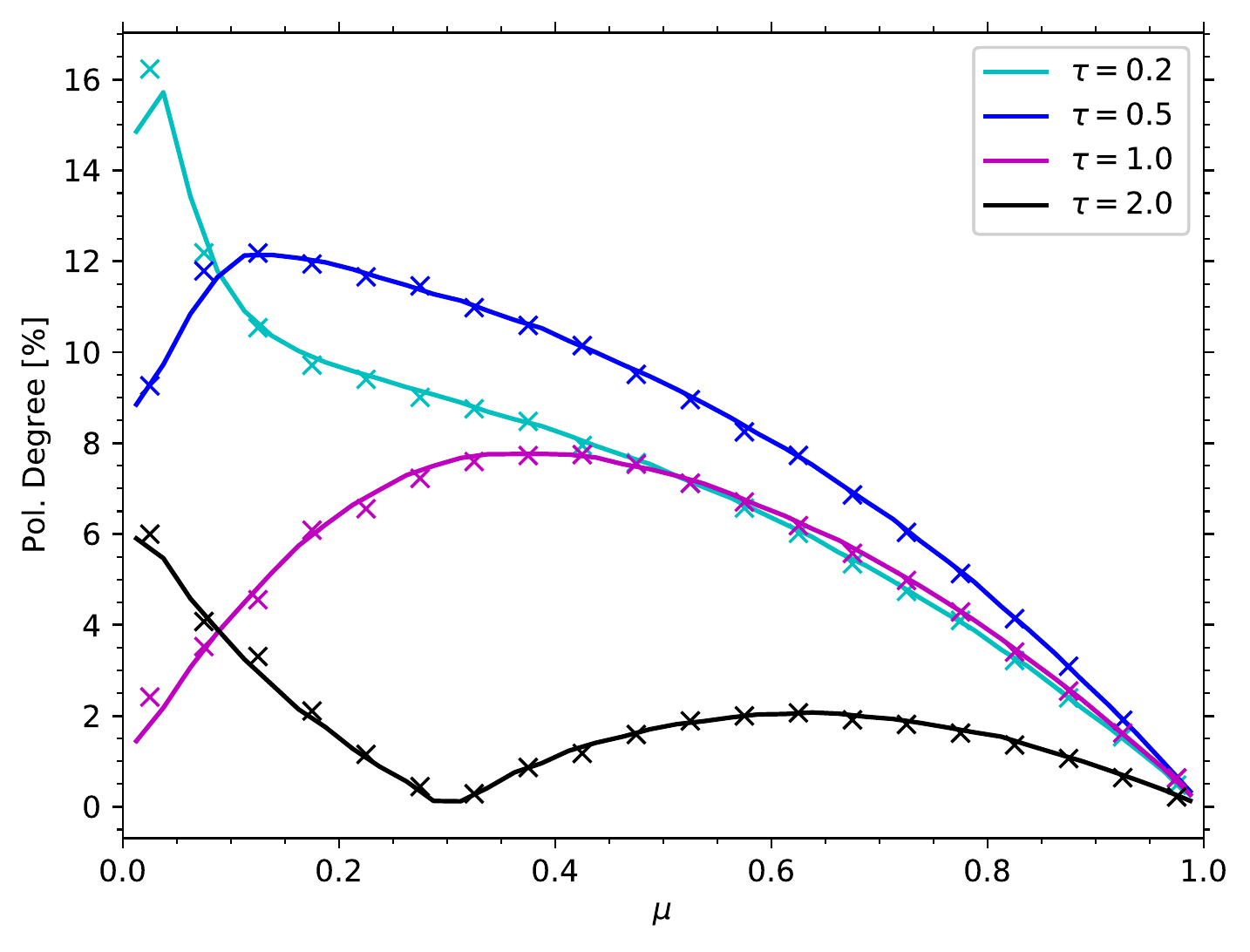}
 \caption{The angular dependence of the polarisation degree of the radiation
 from a pure scattering disc atmosphere.
 Results of \codename{} and \textsc{Stokes} are plotted in crosses and solid lines,
 respectively. The discs with different optical depths are plotted in different colors.\label{fig:pol_plane}}
\end{figure}

We calculate the angular dependence of the polarisation degree of the radiation 
emerging from the surface of a pure scattering disc atmosphere 
and compare the result with \textsc{Stokes} \citep{goosmann_modeling_2007}.
The optical depth is defined as $\tau_{\rm T} = n_e \sigma_{T} h$, 
where $h$ is the half-thickness of the disc.
The primary photons are isotropic and uniformly located on the disc midplane. To compare with \textsc{Stokes}, we
assume Thomson cross section and stationary electrons. Instead of using Eq.~\ref{eq:knstokes}, we
utilise the Rayleigh matrix method to calculate the Stokes parameters of the scattered photons following \citet{schnittman_monte_2013}.
The results are presented in Fig.~\ref{fig:pol_plane}, in which good agreement is seen for various optical depths.


\section{Results}
In this section we showcase a few spectra of disc-corona systems calculated by \codename{}, leaving a systematic study to a future work.
Note that the polarisation option is switched off for all calculations included in this section.
\subsection{Stationary spherical coronae}
In this section we investigate the spectral properties of disc-corona systems in which the coronae are spherical and stationary (i.e., the angular velocity
of the fluid is the same with a ZAMO observer).
\subsubsection{Dependence on the optical depth}
\label{sec:sphere_tau}
\begin{figure}
 \includegraphics[width=\columnwidth]{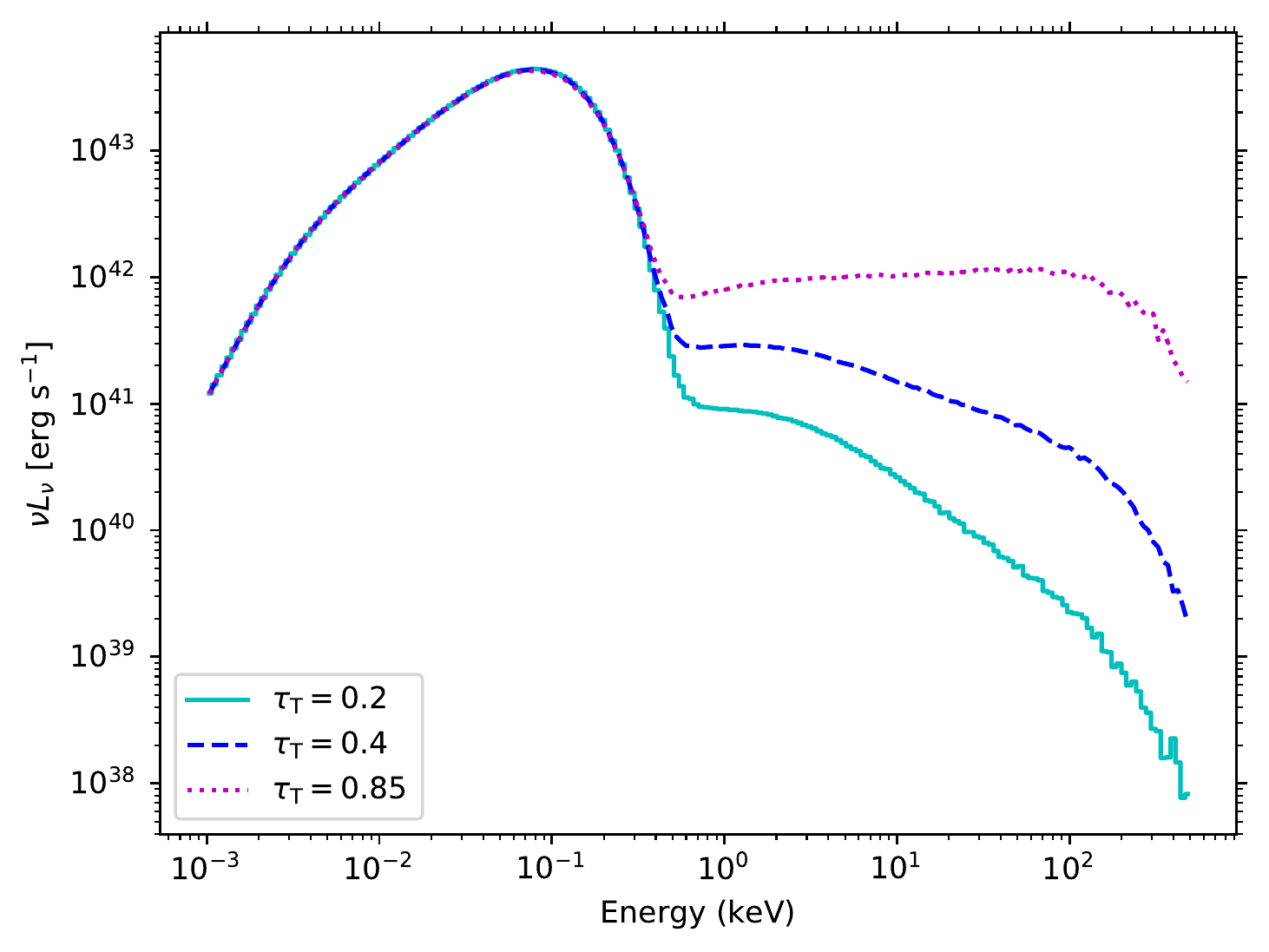}
 \caption{Spectra of disc-corona systems with coronae of various optical depths $\tau_{\rm T}$, as calculated with \codename. The observer
 is located at an inclination of $30^\circ$. Results for different
 optical depths are plotted in different colors and line styles, as indicated in the plot.
 The spherical corona is located $10~\rm GM~c^{-2}$ above the disc, has a radius of $4~\rm GM~ c^{-2}$, and 
 has an electron temperature of $100~\rm keV$.
 The other parameters are: $a=0.998$, $M=10^7~\rm M_\odot$, and $\dot{M} = 4.32\times10^{23}~\rm g~s^{-1}$.\label{fig:sphere_tau}}
\end{figure}

In Fig.~\ref{fig:sphere_tau} we present the spectra of disc-coronae systems consisting coronae of different optical depths.
The disc is around a Kerr black hole with spin $a=0.998$ and mass $M=10^7~\rm M_\odot$.
We assume that the disc is extending down to the innermost stable orbit and there is zero torque at the inner edge of the disc.
The radiative efficiency for a standard disc around a spin 0.998 black hole is expected to be $\sim0.32$.
The outer boundary of the disc is taken to be $1000~\rm GM~c^{-2}$.
The mass accretion rate $\dot{M} = 4.32\times10^{23}~\rm g~s^{-1}$, 
corresponding to a bolometric luminosity of $1.24\times10^{44}~\rm erg~s^{-1}$, $\sim10\%$ the Eddington luminosity.
We take the color correction $f_{\rm col}=2.4$ that is expected for AGN discs \citep{ross_spectra_1992,done_intrinsic_2012}.
The spherical corona is located $10~\rm GM/c^2$ above the disc, has a radius of $R_c = 4~\rm GM/c^2$, 
and has an electron temperature of $100~\rm keV$.
The observer is located at an inclination of $30^\circ$.
We calculate the spectra for different Thomson optical depths $\tau_{\rm T}=0.2, 0.4, 0.85$, the same with \citetalias{dovciak_minimum_2016}, 
while $\tau_{\rm T}\equiv n_e \sigma_{\rm T} R_c$ and $R_c$ is the corona radius.
As the optical depth increases, the thermal spectrum barely varies,
while the non-thermal spectrum hardens and brightens as expected.

\subsubsection{Dependence on the observer's inclination}
\label{sec:incl}
\begin{figure}
 \includegraphics[width=\columnwidth]{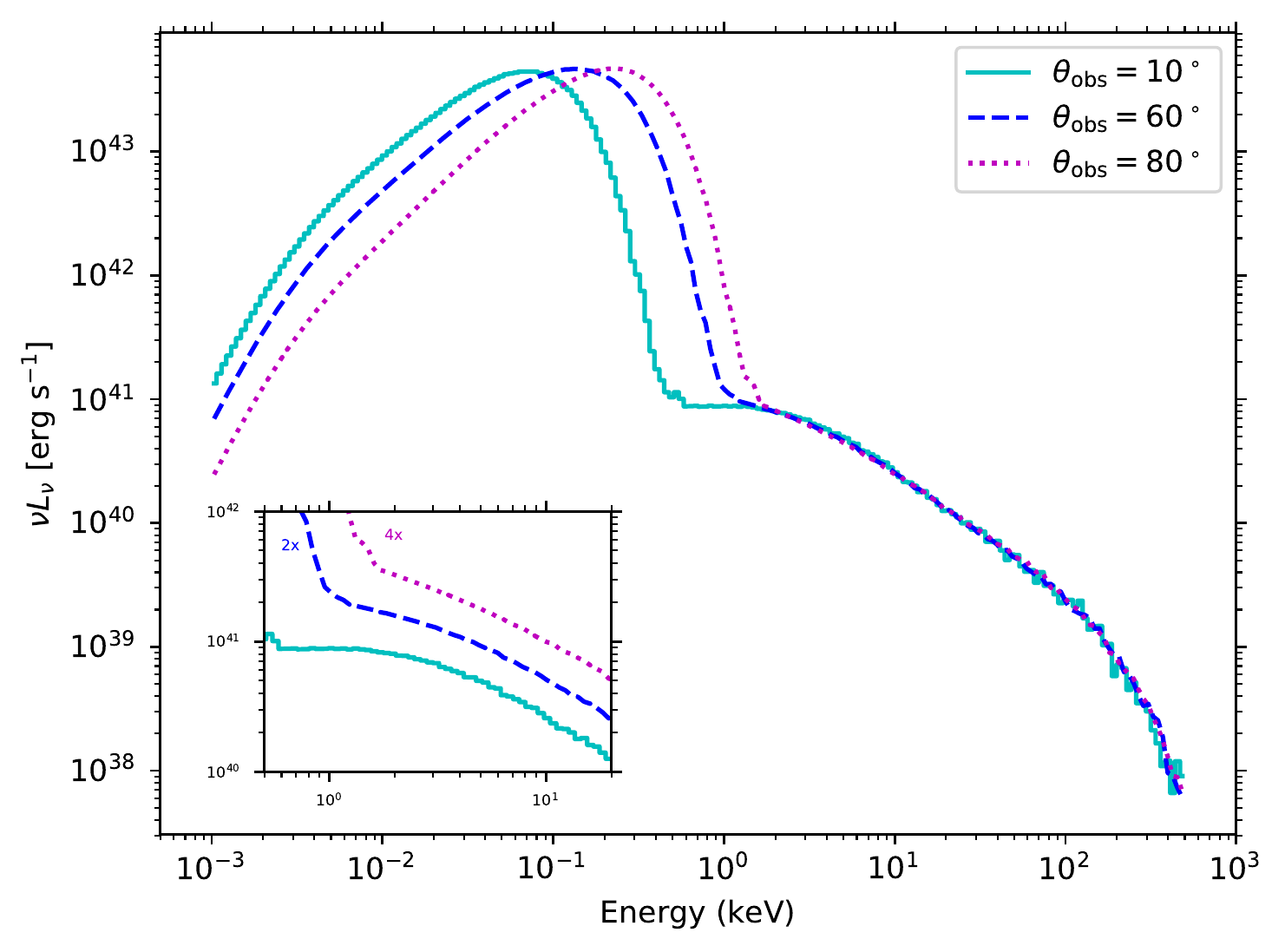}
 \caption{Same as Fig.~\ref{fig:sphere_tau}, but for different observer's inclinations as indicated in the plot and $\tau_{T}=0.2$.
 The inset shows the zoom-in plot of the spectra in the range of 0.5--20 keV, where the spectra for inclinations
 of $60^\circ$ and $80^\circ$ are shifted along the vertical direction for clarity.
 \label{fig:sphere_incl}}
\end{figure}

In Fig.~\ref{fig:sphere_incl} we present the spectra of a disc-corona system as seen by observers at various inclinations. The parameters
are the same as in Section~\ref{sec:sphere_tau}, except that $\tau_{\rm T}=0.2$ and various inclinations are assumed.
To highlight the distinction we zoom-in the spectra
between 0.5 and 20 keV in the inset plot, in which the spectra for inclinations of $60^\circ$ 
and $80^\circ$ are multiplied by factors of $2$ and $4$, respectively, for clarity.
The spectra depend sensitively on the observer's inclination. The spectrum at the inclination of $10^\circ$ flattens towards low energy, and
seems to be harder than that observed at larger inclinations below $\sim5~\rm keV$.

We measure the photon indices between 2 and 10 keV by fitting the spectra with 
powerlaw model using the least square method. The indices are 2.69,
2.70, and 2.73 for inclinations of $10^\circ$, $60^\circ$, and $80^\circ$, respectively. 
For comparison we calculate the spectrum of a spherical plasma cloud Comptonising central isotropic radiation,
the same geometry as in Sec.~\ref{sec:srsphere}. In this case we take $\tau_{\rm T}=0.2$, and $T_e=100~\rm keV$. We also take the temperature
of the thermal radiation to be $0.026~\rm keV$, the temperature of the thermal radiation received by a lamp-post corona with 
a height of $10~\rm GM~c^{-2}$.
We measure the photon index in the isotropic case and find it to be 2.88 between 2 and 10 keV, softer
than all spectra presented in Fig.~\ref{fig:sphere_incl}.

\begin{figure}
 \includegraphics[width=\columnwidth]{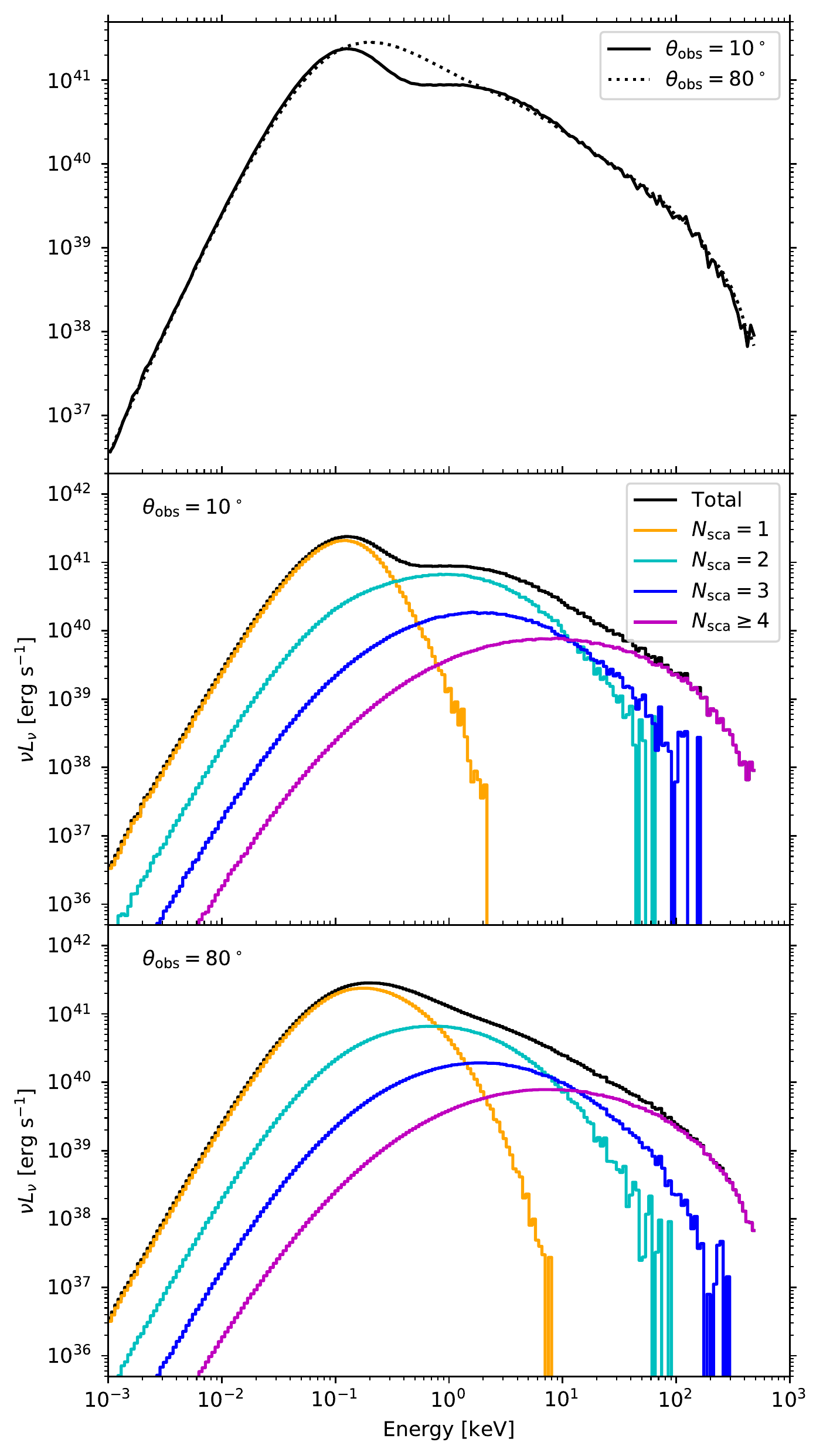}
 \caption{Top panel: the \textit{Comptonised} spectra as observed by observers at different inclinations. Middle and bottom panels: the spectra of individual spectral components that contain photons having experienced
 different numbers of scattering. The spectra in the middle and bottom panels correspond to the $10^\circ$ and $80^\circ$ spectra in Fig.~\ref{fig:sphere_incl}, 
 respectively.
 \label{fig:sphere_2nsca}}
\end{figure}

%
The angular dependence is more clearly seen in Fig.~\ref{fig:sphere_2nsca} 
where we present the \textit{Comptonised} spectra as well as contribution of photons with different 
numbers of scattering. Between $\sim0.1$ and $2$ keV the spectrum observed at higher inclination is softer and more luminous.
By inspecting the contributions of different scattering orders (the middle and bottom panels), we find out that this is mainly
caused by the difference in the scattering spectrum of the first order.

The angular dependence is due to anisotropic illumination of the seed photons.
For isotropic electrons scattering anisotropic photons, 
the first scattering spectrum would be highly anisotropic,
with more scattering power backward to the direction of the seed photons than forward;
while the high order spectra are expected to be isotropic
\citep[e.g.,][]{ghisellini_anisotropic_1991,haardt_two-phase_1991,haardt_anisotropic_1993}.
Therefore for the geometry we assume, where the seed photons are emitted by the thin disc below the corona,
more powerful first scattering spectrum at larger inclination is expected, consistent with what we see in Fig.~\ref{fig:sphere_2nsca}.
This indicates that in the corona frame the radiation illuminating the underlying disc is more 
luminous than the radiation that arrives at an
observer at infinity. This is contrary to the isotropic corona assumption
usually taken in modelling the reflection spectrum from the disc illuminated by a lamp-post corona.
We will further discuss this topic in Sec.~\ref{sec:discussion}.

\subsubsection{Dependence on the size of the corona}
\begin{figure}
 \includegraphics[width=\columnwidth]{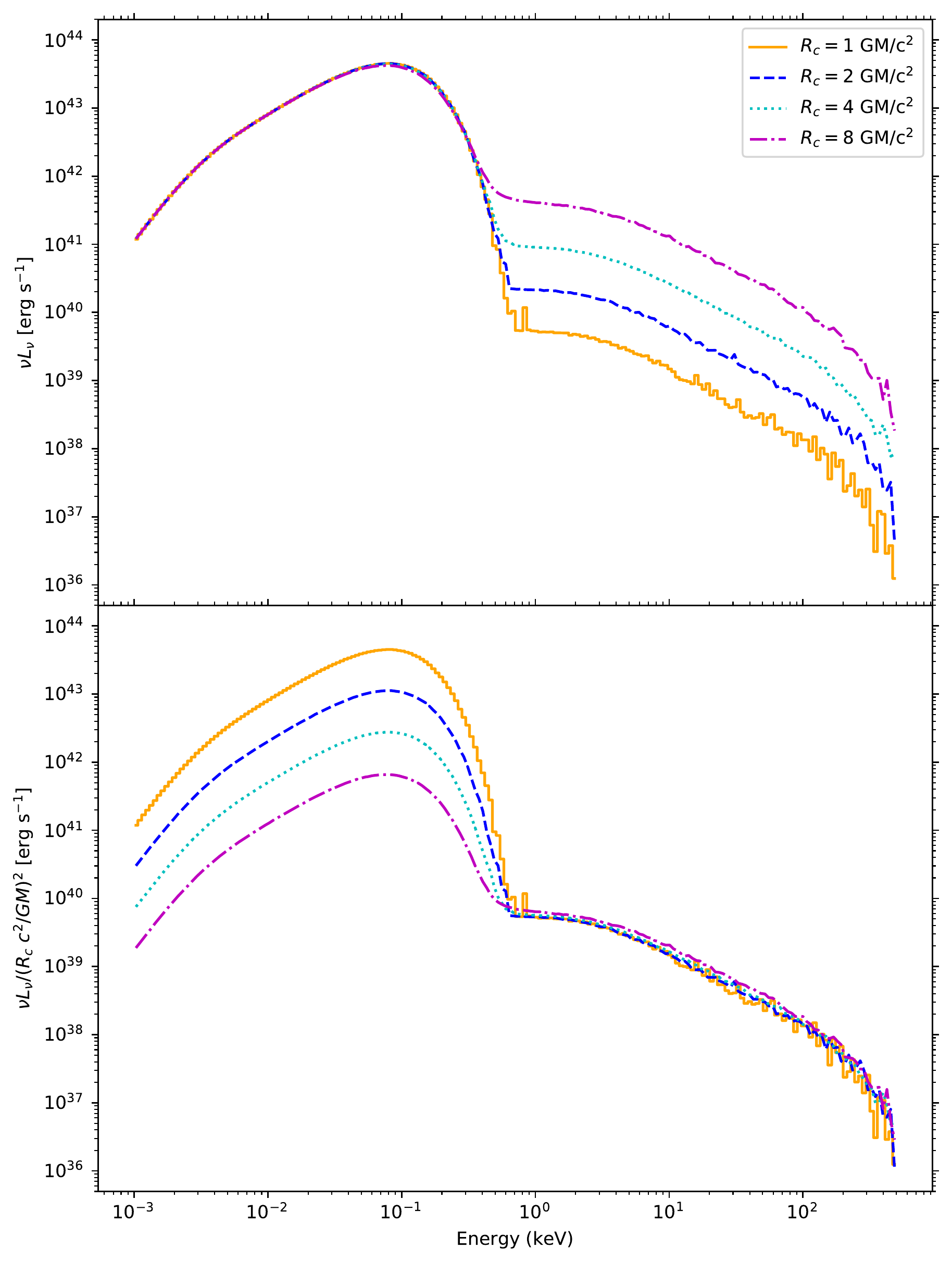}
 \caption{Upper panel: same as Fig.~\ref{fig:sphere_tau}, but for different corona radii as indicated in the plot and $\tau_{T}=0.2$.
 The observer's inclination is $30^\circ$.
 Lower panel: spectra normalised by $R_c^2$, where $R_c$ is the corona radius in $\rm GM~c^{-2}$.\label{fig:sphere_size}}
\end{figure}

In Fig.~\ref{fig:sphere_size} we present the spectra of disc-corona systems with different corona radii.
The parameters are the same as in Section~\ref{sec:sphere_tau}, but with various sizes of the corona and $\tau_{\rm T} = 0.2$.
The most prominent effect is that the non-thermal luminosity
becomes brighter as $R_c$ increases. 
In the lower panel of Fig.~\ref{fig:sphere_size} we show the spectra normalized by $R_c^2$. For corona emission the normalised spectra
agree with each other quite well, indicating that the Comptonised spectra scale with $R_c^2$.


For each spectrum in Fig.~\ref{fig:sphere_size} 
we estimate the size of the corona using the \citetalias{dovciak_minimum_2016} method with the measured photon index and luminosity.
For corona sizes of 1, 2, 4, and 8 $\rm GM~c^{-2}$, the \citetalias{dovciak_minimum_2016}
method gives estimates of 0.63, 1.30, 2.55, and 5.00 $\rm GM~c^{-2}$ on the corona radius, respectively, 
$\sim$ half the input sizes. The reason is that in \textsc{nthcomp} the emergent spectrum is obtained under the
assumption that the seed photons are uniformly distributed and isotropic. As shown in Section~\ref{sec:incl}, with the same optical depth the
spectrum in the isotropic case is softer than the spectrum observed at low inclination if the seed photons are illuminated anisotropically. 
Hence the \citetalias{dovciak_minimum_2016} method, which assumes isotropic seed photons, requires a larger than assumed
optical depth to explain the spectrum. This subsequently leads to an overestimate of the luminosity and
an underestimate of the size of the corona.

We also calculate the photon index and luminosity as measured by observers at inclinations of $10^\circ$ and 
$60^\circ$ and estimate the size of the corona using the \citetalias{dovciak_minimum_2016} method.
For an inclination of $10^\circ$, the \citetalias{dovciak_minimum_2016} method gives estimate of 0.64, 1.31, 2.57, and 5.08
$\rm GM~c^{-2}$ on the corona radius for the input corona sizes of 1, 2, 4, and 8 $\rm GM~c^{-2}$, respectively; whereas for an
inclination of $60^\circ$, the estimated sizes are 0.64, 1.26, 2.55, and 5.14 $\rm GM~c^{-2}$, respectively.
For both inclinations the estimated sizes are close
with the estimated sizes when the observer's inclination is $30^\circ$.

\subsection{Rotating spherical coronae}
\label{sec:rotsphere}
In this section we investigate the effect of corona rotation on the emergent spectrum. The set-up is the same as in Sec.~\ref{sec:incl}, except that
the corona is not stationary, but rotating about the black hole. 
To assess the effect of rotation, we make a simple assumption on corona motion that
everywhere in the corona we take the linear circular velocity to be $c/2$ as measured by a ZAMO observer.
This is approximately the rotation velocity of the disc fluid at the inner edge of the thin disc around a black hole with spin $a=0.998$.
Calculations based on more physically motivated corona models will be carried out in a future work.

\begin{figure}
 \includegraphics[width=\columnwidth]{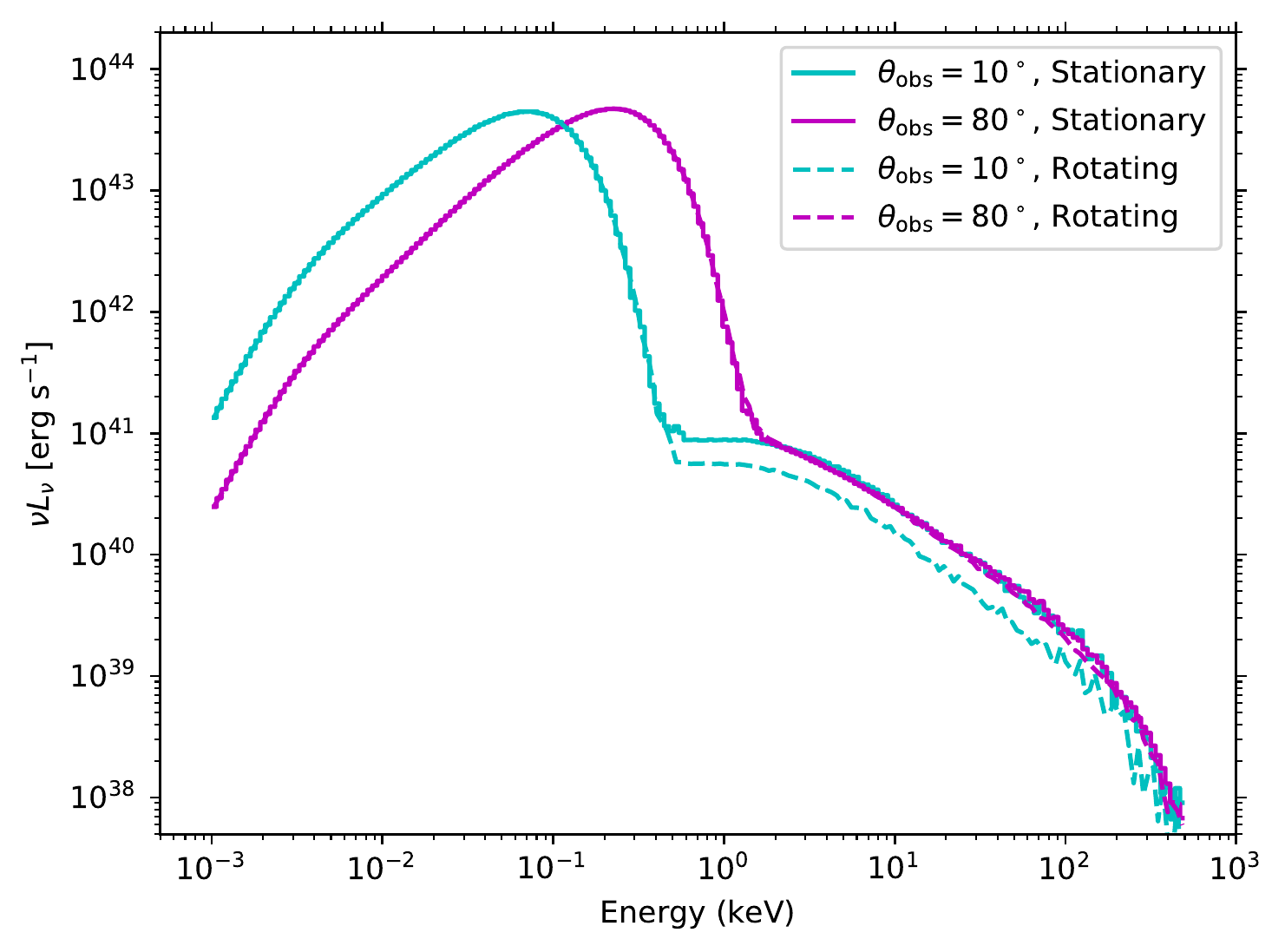}
 \caption{The spectra of a disc-corona system that contains a rotating spherical corona as seen by observers at
 different inclinations, in dashed lines. Everywhere in the corona the fluid is rotating with a linear velocity of $c/2$ as
 measured by a ZAMO observer. The other parameters are the same as Fig.~\ref{fig:sphere_incl}. For comparison we also plot the spectra 
 for stationary corona in solid lines.
 \label{fig:sphere_rotation}}
\end{figure}

In Fig.~\ref{fig:sphere_rotation} we present the inclination-dependent spectra (dashed lines) and compare them with stationary coronae (solid lines).
While at a large inclination angle the rotation has little effect,
at a low inclination the non-thermal emission is significantly less luminous (by a 
factor of 1.5 compared with the high inclination spectrum) due to
beaming effect. This suggests that compared with stationary spherical corona,
for rotating spherical corona one would further underestimate the the size of the corona with the
\citetalias{dovciak_minimum_2016} method if the observer is located at a low inclination.

\subsection{Co-rotating slab coronae}
In this section we present the spectra for different disc-corona systems in which the coronae are of slab geometry above the disc.
We assume that the coronae are co-rotating with the underlying Keplerian thin discs.
\subsubsection{Dependence on the observer's inclination}
\label{sec:slabincl}
\begin{figure}
 \includegraphics[width=\columnwidth]{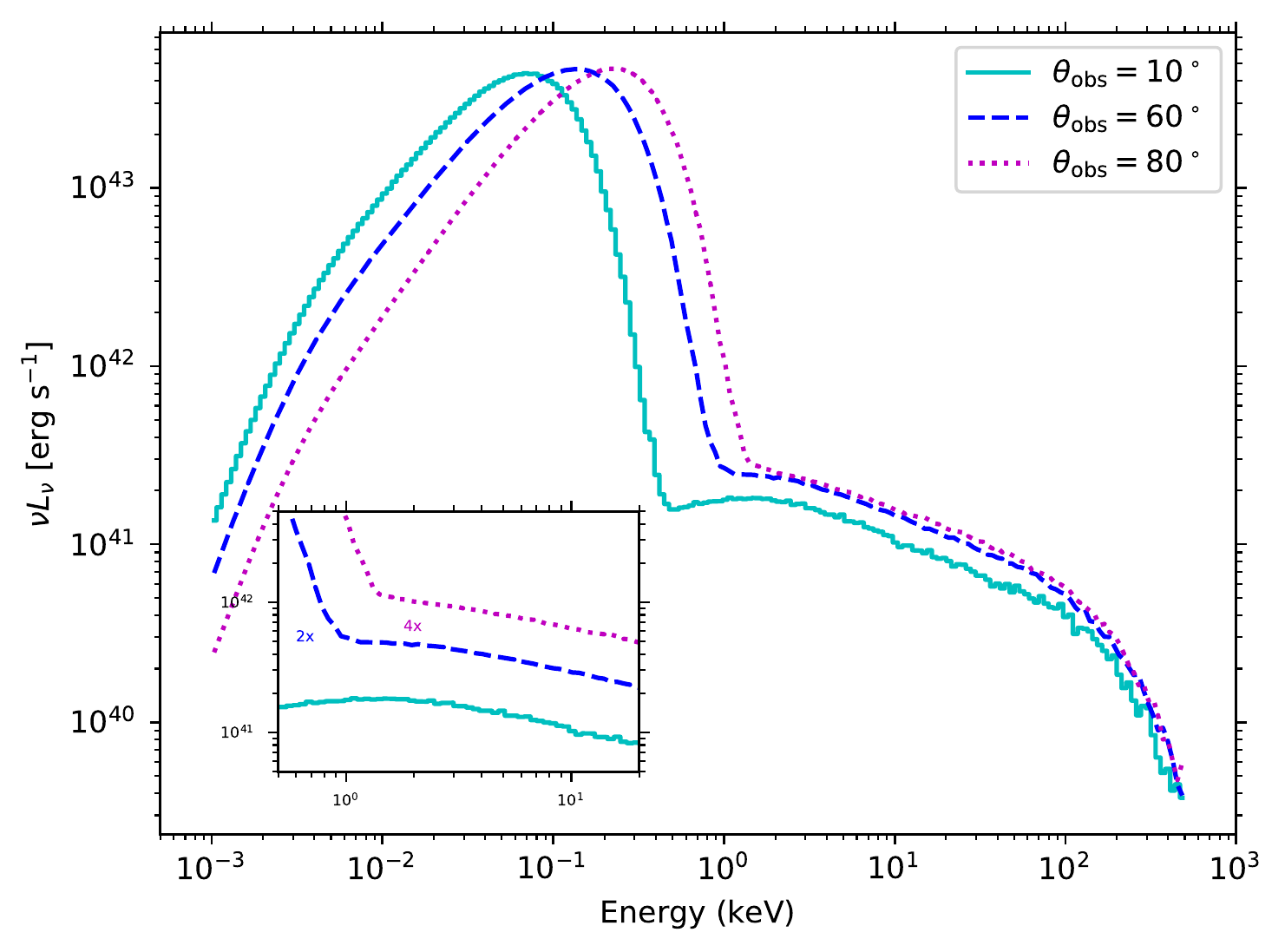}
 \caption{Spectra of a disc-corona system as seen by observers at various inclinations. 
The co-rotating slab corona is located between $9$ and $11~\rm GM~c^{-2}$ above the disc.
The radius of the corona is $4~\rm GM~c^{-2}$. The Thomson optical depth along the vertical direction is 0.2.
 Other parameters are the same as in Fig.~\ref{fig:sphere_tau}.\label{fig:slab_incl}}
\end{figure}

In Fig.~\ref{fig:slab_incl} we present the spectra of a disc-corona system as seen by observers at various inclinations. 
The co-rotating slab corona has a height of $10~\rm GM~c^{-2}$, a thickness of $2~\rm GM~c^{-2}$, 
and a radius of $4~\rm GM~c^{-2}$.
Its temperature is 100 keV, and its optical depth along the vertical direction of $\tau_{\rm T} = n_e \sigma_{\rm T} h = 0.2$, where
$h=1~\rm GM~c^{-2}$ is the half-thickness of the disc. The other parameters are the same as in Section~\ref{sec:sphere_tau}.
The observers at lower inclinations see less luminous X-ray emission, the same with rotating spherical corona (Sec.~\ref{sec:rotsphere}).

There is also difference in the spectral shape below $\sim5~\rm keV$.
To highlight the distinction we plot the zoom-in spectra
between 0.5 and 20 keV in the inset plot, in which the spectra for inclinations of $60^\circ$ 
and $80^\circ$ are multiplied by a factor of $2$ and $4$, respectively, for clarity.
Similarly with spherical coronae, the low-inclination spectrum 
below $\sim5~\rm keV$ seems to be harder than that observed at larger inclinations, and flattens towards low energy.

\subsubsection{Dependence on the size of the corona}
\begin{figure}
 \includegraphics[width=\columnwidth]{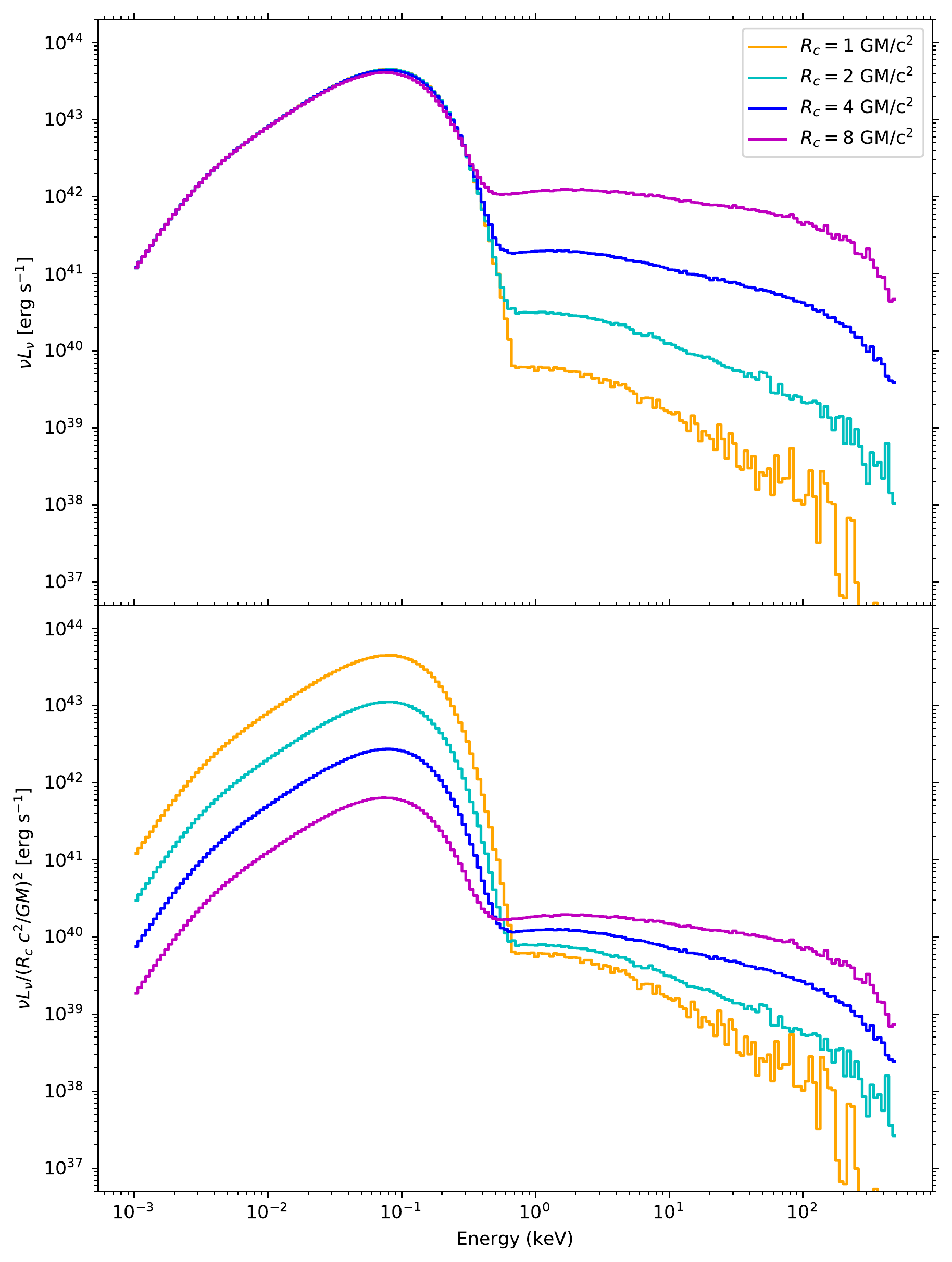}
 \caption{Upper panel: the same with Fig.~\ref{fig:slab_incl}, but for coronae of different sizes. The observer is located at an
 inclination of $30^\circ$. Spectra of different corona sizes are plotted in different colors, as indicated in the plot.
 Lower panel: the spectra are divided by $R_c^2$, where $R_c$ is the corona radius in $\rm GM~c^{-2}$.\label{fig:slab_size}}
\end{figure}

In Fig.~\ref{fig:slab_size} we present the energy spectra of disc-corona systems consisting of slab coronae of different sizes.
The parameters are the same as in Section~\ref{sec:slabincl}, but with different radii and the observer is located at an inclination of $30^\circ$.
Similarly with Fig.~\ref{fig:sphere_size}, in the lower panel of Fig.~\ref{fig:slab_size} we plot the spectra normalised by $R_c^2$.
Unlike spherical coronae, the spectra of slab coronae do not simply scale with $R_c^2$, but become harder and more luminous as the radius increases, due to
increased optical depth along the horizontal direction.

\section{Discussion}
\label{sec:discussion}
As seen in Fig.~\ref{fig:sphere_incl}, the corona emission is not isotropic even for a stationary spherical corona.
Owing to anisotropic illumination of the seed photons,
the Comptonised spectrum is more powerful toward the direction of the seed photons, indicating that 
for the coronae in AGNs above the disc, in the corona rest frame there would be more powerful Comptonised radiation 
striking the underlying disc than to an observer at
infinity. This is contrary to the assumption of an isotropic corona, which is usually taken in modelling and
interpreting the reflection spectrum, with a few exceptions \citep[e.g.,][]{henri_anisotropic_1997,petrucci_anisotropic_1997,malzac_anisotropic_1998}.

\begin{figure}
\includegraphics[width=\columnwidth]{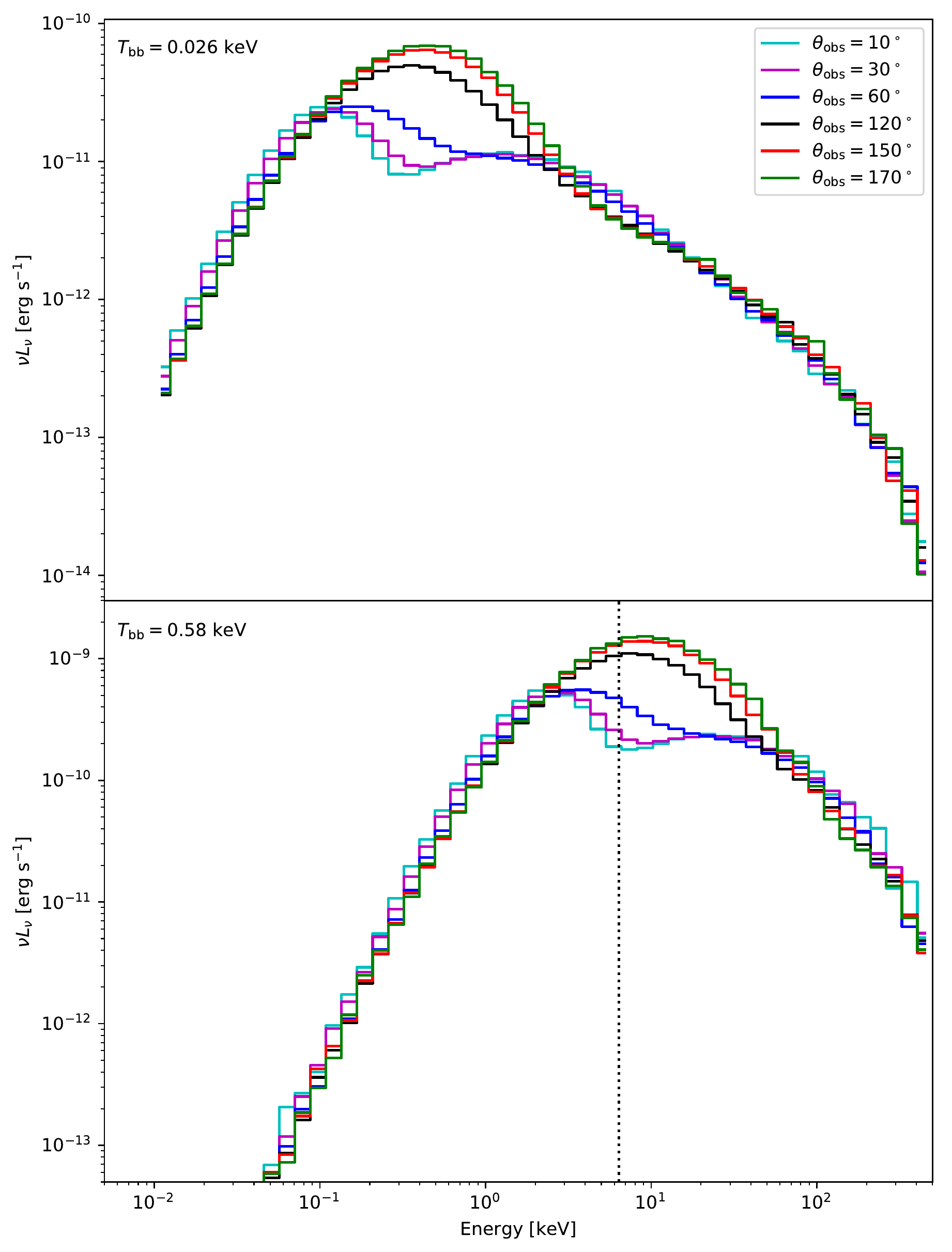}
 \caption{Upper panel: the spectra of a spherical plasma cloud Comptonising thermal seed photons, as seen by observers at different inclinations.
 The seed photons are moving along a single direction and the inclination is defined as the angle made by the seed photon direction and
 the line of sight to the observer.
 Spectra observed at different inclinations are plotted in different colors and line styles, as indicated in the plot.
 The plasma has a temperature of $100~\rm keV$ and a Thomson optical depth of $0.2$. The temperature of the seed 
 photons is $0.026~\rm keV$.
 The seed photon injection rate is $1~\rm counts~s^{-1}$. Lower panel: same as the upper panel,
 but for seed photon temperature of $0.58~\rm keV$.
 We also indicate the energy of neutral iron K edge at 7.1 $~\rm keV$ with a vertical dashed line.
 \label{fig:srsphere_incl}}
\end{figure}

To investigate the difference between the illuminating and direct emission, we calculate the angular dependent spectra for spherical plasma Comptonising
highly anisotropic seed photons in which the seed photons are moving along a single direction.
We take the electron temperature $T_e=100 ~\rm keV$ and the optical depth $\tau_{\rm T}=0.2$. 
To assess the effect in both AGNs and black hole X-ray binaries (BHXRBs),
we perform calculations for two different seed photon temperatures of $0.026~\rm keV$ and $0.58~\rm keV$,
the temperatures of seed photons as received by lamp-post coronae with a height of $10~\rm GM~c^{-2}$ above black holes accreting at $10\%$ Eddington
rate with masses of $10^7~\rm M_\odot$ and $10~\rm M_\odot$, respectively.
In Fig.~\ref{fig:srsphere_incl} we present the spectra for observers at various inclinations, where the inclination
is defined to be the angle between the line of sight and the seed photon direction. 
Hence inclinations of $10^\circ$, $30^\circ$, and $60^\circ$ correspond to distant observers; whereas
inclinations of $120^\circ$, $150^\circ$, and $170^\circ$ represent observers on the underlying disc.

\begin{figure}
\includegraphics[width=\columnwidth]{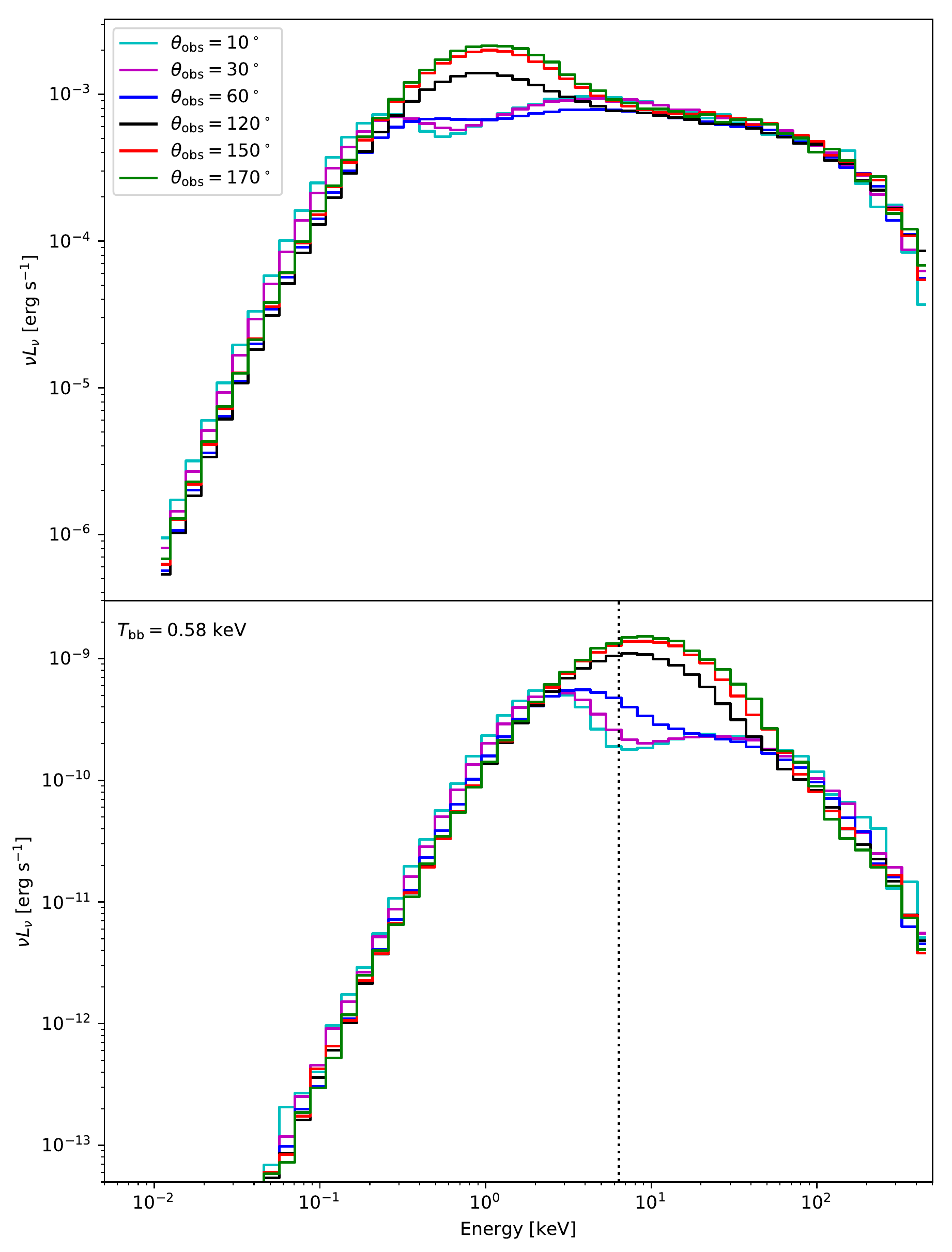}
 \caption{The same as Fig.~\ref{fig:srsphere_incl}, but for a slab plasma cloud. The Thomson optical depth of the cloud
 $\tau_{\rm T} \equiv n_e \sigma_{\rm T} h/2=0.2$, where $h$ is the thickness of the cloud. The radius of the cloud along the horizontal
 direction is 5 times its thickness.\label{fig:srslab_incl}}
\end{figure}

In the upper panel of Fig.~\ref{fig:srsphere_incl} we present the results for the AGN case.
Above $\sim3~\rm keV$ the corona emission is more or less isotropic.
As the illuminating radiation interacts with the cold disc mainly via Compton down-scattering and photo-ionisation,
the estimate of the flux density of the reflection emission above $\sim3~\rm keV$ will not be affected much by assuming an isotropic corona.
However, there is large contrast between the spectra of the radiation received by observers on the disc (with inclinations of $120^\circ$, 
$150^\circ$, and $170^\circ$) and that by
distant observers (with inclinations of $10^\circ$, $30^\circ$, and $60^\circ$) in the range of $\sim0.1-3~\rm keV$.
A common practice in modelling the reflection spectrum is to take the photon index of the observed X-ray continuum
to be the photon index of the illuminating radiation.
In the range of 2--10 keV, the photon indices are $2.68$, $2.64$, and $2.65$ for inclinations of $10^\circ$, $30^\circ$, and
$60^\circ$, respectively; while the observers on the disc see softer spectra, with indices of $2.86$, $3.12$, and $3.27$ for inclinations
of $120^\circ$, $150^\circ$, and $170^\circ$, respectively.
Therefore by assuming the corona radiation to be isotropic in the corona rest frame, one may
underestimate the photon index of the illuminating radiation.
This can be eased if the photon index is measured at a
harder energy band. For example, in the energy band of 10--79 keV, the photon indices differ no more than 0.2 among
different inclinations.

Many AGN spectra contain ``soft excess'' below 1 keV compared with the $2-10~\rm keV$ continuum
\citep[e.g.,][]{turner_variability_1988}.
One popular scenario for the origin of the soft excess in AGNs is that it is due to hard corona emission
reflecting off ionised disc material
\citep[e.g.,][]{crummy_explanation_2006}.
Unlike the reflection continuum, the flux of the soft excess would be underestimated due to anisotropy of 
the corona emission below 3 keV.

The spectra for the BHXRB case are presented in the lower panel of Fig.~\ref{fig:srsphere_incl}.
Similarly with the AGN case, the spectra to the distant observers
are harder and less luminous, and flatten towards low energy. 
The photon index of the illuminating emission would be underestimated if one assumes that the corona emission is isotropic in the corona
rest frame. Even if the photon index is measured in a harder band, e.g., 10--79 keV, 
the photon indices are $2.10$, $2.15$, and $2.36$ for observers at inclinations of $10^\circ$, $30^\circ$, and $60^\circ$, 
respectively; while much softer spectra are seen by observers on the disc, with photon indices of $3.24$, $3.23$, and $3.21$
for observers at inclinations of $120^\circ$, $150^\circ$, and $170^\circ$, respectively.

For an isotropic lamp-post corona,
the reflection fraction is expected to be close to unity in the Newtonian case \citep{dauser_role_2014}. 
While in the case of fast spinning black hole,
the gravitational field focus the radiation towards the disc and reduce the intensity received by an observer at infinity, 
thus a larger value of reflection fraction is expected, especially for a 
compact corona close to the black hole \citep[e.g.,][]{miniutti_light_2004,dauser_role_2014}.
However, for the BHXRB case, the corona emission is anisotropic in the range of $\sim3-60~\rm keV$, a harder band than AGNs.
In this case the estimate of the reflection continuum flux density will also be affected. 
We calculate the luminosity above 1 keV, and find that the luminosity measured by observers on the disc is in general
larger than the luminosity measured by distant observers. For instance, at $170^\circ$ the luminosity is $\sim3$ times that at $10^\circ$.
In this case a reflection fraction larger than unity can be observed even without a fast spinning black hole.

For the BHXRB case
the largest contrast in the spectra is around $10~\rm keV$.
As the photo-ionisation cross section decreases rapidly with energy above the absorption edge, 
the photons with energy just above iron K edge are most
essential for the production of fluorescent iron K photons
\citep{george_x-ray_1991}. 
Hence the anisotropy of the corona emission has an even larger effect in estimating the 
flux of the fluorescent iron K line than the reflection continuum.


While the analysis above hints us qualitatively the consequence of assuming isotropic corona, to quantitatively assess the effect 
more issues have to be taken into account. The angular distribution of the seed photons is more complicated.
The gravitational field of the black hole will play an important role in modifying the energies
and trajectories of the illuminating photons, especially when the black hole has large spin.
For an observer at infinity the direct emission is contaminated by the reflection emission, and to
assess the effect one needs to take into account the reflected photon arriving at infinity as well.
Such kind of calculation can be done with \codename{} with a reflection model,
by calculating the reflection spectrum for photons arriving at the disc and propagating the reflected
photons in the same fashion as the thermal seed photons.
However this is out of the scope of this paper and will be carried out in a future work.

We also investigate the angular dependent spectra for slab plasma clouds and present the results in Fig.~\ref{fig:srslab_incl}. 
It is obvious that spectra depend on the observer's inclination in a similar fashion as the spherical case, therefore the
conclusion we draw based on the spherical cloud case should apply to slab geometry as well.

\section{Summary}
The size of the X-ray corona in AGNs is still not well constrained.
To date the most promising constraint comes from analysis of strongly lensed quasars, but the paucity of them
limits the application of this method. \citetalias{dovciak_minimum_2016} developed a method to measure the size of the corona
with simultaneous X-ray and UV observations,
while the corona spectrum was calculated assuming a lamp-post geometry. To perform more self-consistent calculations
of the corona spectra, we
develop \codename{}, a Monte Carlo radiative transfer code that is dedicated to the Comptonisation process in the Kerr spacetime.
We include all general relativistic effects
and assume Klein-Nishina scattering cross section. We compare the results by \codename{} with that by previous codes and find them to be consistent.

We calculate spectra of disc-corona systems in AGNs that consist of Novikov-Thorne discs on the equatorial plane
and optically-thin coronae above the geometrically-thin discs.
For stationary spherical coronae, owing to anisotropic illumination of the seed photons, 
the corona emission is inclination dependent, with observers at lower inclinations seeing harder and less luminous corona emission.
The non-thermal emission scales with the corona radius square. We find that the sizes estimated with the
\citetalias{dovciak_minimum_2016} method
are around half the assumed size. The reason for this discrepancy is that \citetalias{dovciak_minimum_2016} 
assumes isotropic corona emission which is softer than the spectrum
observed at low inclinations at the same optical depth.
For spherical corona that is rotating about the symmetric axis, at a large inclination the spectrum is more or less the
same as stationary spherical corona, while at low inclination the observed spectrum is less luminous than the stationary case due to
beaming effect. For co-rotating slab coronae,
the spectra depend on the observer's inclination in a similar fashion as spherical coronae.
However, unlike spherical spherical coronae, as the size increases the spectrum also hardens.
A thorough study of corona emission will be carried out in a future work.

We discuss the implication of anisotropic corona emission for modelling and interpreting the illuminating spectrum. 
For AGNs this would lead to an inaccurate estimate of the spectral
shape of the illuminating corona emission as well as an underestimate of the flux of 
the illuminating radiation that is accounting 
for the soft excess. For BHXRBs, this would also lead to underestimated reflection fraction and iron K flux.

\acknowledgments
We thank the anonymous referee for his/her careful reading of the manuscript and useful comments and suggestions.
The authors thank Giorgio Matt for valuable comments, and Iossif Papadakis, Jason Dexter for useful discussion.
The authors acknowledge financial support provided by Czech Science Foundation grant 17-02430S. 
This work is also supported by the project RVO:67985815. This research makes use of
\textsc{matplotlib} \citep{hunter_matplotlib_2007},
a Python 2D plotting library which produces publication quality figures.

\appendix
\section{Tetrad}
\label{sec:tetrad}
The method for obtaining an orthonormal tetrad attached to an observer with arbitrary four-velocity can be found in
\citet{sadowski_spinning_2011}.
In the following we give the expressions of the tetrads in two specific cases: 
1) the tetrad attached to a ZAMO observer; 2) the tetrad
attached to an observer moving in the azimuthal direction.

\subsection{ZAMO tetrad}
The four-velocity of a ZAMO observer is
\begin{equation}
 \boldsymbol{U} = \left[U^t,\ 0,\ 0,\ -\frac{g_{t\phi}}{g_{\phi\phi}}U^t\right],
 \end{equation}
 where
\begin{equation}
 U^t = \sqrt{\frac{g_{\phi\phi}}{g_{t\phi}^2 - g_{tt}g_{\phi\phi}}}.
\end{equation}
The orthonormal tetrad attached to a ZAMO observer is
\begin{eqnarray}
 \boldsymbol{e}_{(t)} &=& \boldsymbol{U},\\
 \boldsymbol{e}_{(r)} &=& \left[0,\ \frac{1}{\sqrt{g_{rr}}},\ 0,\ 0\right],\\
 \boldsymbol{e}_{(\theta)} &=& \left[0,\ 0,\ -\frac{1}{\sqrt{g_{\theta\theta}}},\ 0\right],\\
 \boldsymbol{e}_{(\phi)} &=& \left[0,\ 0,\ 0,\ \frac{1}{\sqrt{g_{\phi\phi}}}\right].
\end{eqnarray}
\subsection{Azimuthal tetrad}
For an observer moving in the azimuthal direction with an angular velocity
$\Omega\equiv d\phi/dt$, its four-velocity is
\begin{equation}
 \boldsymbol{U} = \left[U^t,\ 0,\ 0,\ \Omega U^t\right],
\end{equation}
where 
\begin{equation}
 U^t = \sqrt{-\frac{1}{g_{tt} + 2\Omega g_{t\phi} + \Omega^2 g_{\phi\phi}}}.
 \end{equation}
 Denoting 
 \begin{eqnarray}
  k_1 = g_{tt}U^t + g_{t\phi} U^\phi,\\
  k_2 = g_{t\phi}U^t + g_{\phi\phi} U^\phi ,
 \end{eqnarray}
then the time component of $\boldsymbol{e}_{(\phi)}$ can be written as
\begin{eqnarray}
 \boldsymbol{e}_{(\phi)}^t &=& \frac{-sign(k_1)k_2}{\sqrt{(g_{tt}g_{\phi\phi} - 
    g_{t\phi}^2)\left[g_{tt}(U^t)^2 + g_{\phi\phi}(U^\phi)^2 + 2g_{t\phi}U^tU^\phi\right]}},
\end{eqnarray}
and the tetrad is
\begin{eqnarray}
 \boldsymbol{e}_{(t)} &=& \boldsymbol{U},\\
 \boldsymbol{e}_{(r)} &=& \left[0,\ \frac{1}{\sqrt{g_{rr}}},\ 0,\ 0\right],\\
 \boldsymbol{e}_{(\theta)} &=& \left[0,\ 0,\ -\frac{1}{\sqrt{g_{\theta\theta}}},\ 0\right],\\
 \boldsymbol{e}_{(\phi)} &=& \left[\boldsymbol{e}_{(\phi)}^t,\ 0,\ 0, -\frac{k_1}{k_2}\boldsymbol{e}_{(\phi)}^t\right].
\end{eqnarray}

\end{document}